%% file: DecInt_v5p1b.tex
\documentclass[aps,pra,preprint,amsmath,amssymb]{revtex4}
\usepackage{graphicx,epsfig,epsf}
\usepackage{dcolumn}
\usepackage{bm}
\input{diagramsLA}

\begin{document}
\newcommand{\ri}{{\rm i}}
\newcommand{\re}{{\rm e}}
\newcommand{\bx}{{\bf x}}
\newcommand{\bd}{{\bf d}}
\newcommand{\br}{{\bf r}}
\newcommand{\bk}{{\bf k}}
\newcommand{\bE}{{\bf E}}
\newcommand{\bR}{{\bf R}}
\newcommand{\bM}{{\bf M}}
\newcommand{\bn}{{\bf n}}
\newcommand{\bs}{{\bf s}}
\newcommand{\tr}{{\rm tr}}
\newcommand{\tbs}{\tilde{\bf s}}
\newcommand{\rSi}{{\rm Si}}
\newcommand{\beps}{\mbox{\boldmath{$\epsilon$}}}
\newcommand{\bthe}{\mbox{\boldmath{$\theta$}}}
\newcommand{\rg}{{\rm g}}
\newcommand{\xmax}{x_{\rm max}}
\newcommand{\ra}{{\rm a}}
\newcommand{\rx}{{\rm x}}
\newcommand{\rs}{{\rm s}}
\newcommand{\rP}{{\rm P}}
\newcommand{\up}{\uparrow}
\newcommand{\down}{\downarrow}
\newcommand{\hc}{H_{\rm cond}}
\newcommand{\kb}{k_{\rm B}}
\newcommand{\cI}{{\cal I}}
\newcommand{\tit}{\tilde{t}}
\newcommand{\cE}{{\cal E}}
\newcommand{\cC}{{\cal C}}
\newcommand{\Ubs}{U_{\rm BS}}
\newcommand{\qq}{{\bf ???}}
\newcommand*{\etal}{\textit{et al.}}

\newcommand{\be}{\begin{equation}}
\newcommand{\ee}{\end{equation}}
\newcommand{\bfg}{\begin{figure}}
\newcommand{\efg}{\end{figure}}
\newcommand{\bra}{\langle}
\newcommand{\ket}{\rangle}
\newcommand{\Itwo}{\mathbb{1}_2}
\newcommand{\I}{\mathcal{I}}
\newcommand{\al}{\alpha}

\sloppy

\title{Distribution of Interference in the Presence of Decoherence} 
\author{Ludovic Arnaud and Daniel Braun}
\affiliation{Universit\'e de Toulouse, UPS, Laboratoire
de Physique Th\'eorique (IRSAMC), F-31062 Toulouse, France}
\affiliation{CNRS, LPT (IRSAMC), F-31062 Toulouse, France}

\begin{abstract}
We study the statistics of quantum interference for completely
positive maps. We calculate analytically the mean interference and its
second moment for finite dimensional quantum systems interacting with
a simple environment consisting of one or several spins (qudits). The
joint propagation of the entire system is taken as unitary with an
evolution operator drawn from the Circular Unitary Ensemble (CUE).  We
show that the mean interference decays with a power law as function of
the dimension of the Hilbert space of the environment, with a power
that depends on the temperature of the environment. 
\end{abstract}
\maketitle

\section{Introduction}
Quantum information theory predicts increased computational power for
quantum algorithms compared to classical algorithms.  The most
well--known example is Shor's algorithm which factors a
large integer number in a time which grows only polynomially in the
number of digits \cite{Shor94}, whereas no such algorithm is known
classically.  Grover found a quantum algorithm that  allows to  
find an item in an unstructured data base of size $N$ with a number of
queries that 
scales only like $\sqrt{N}$, whereas classically the number of queries
is of order $N$.  Exponential acceleration compared to the best known classical algorithm 
was also predicted for the
shifted character problem  \cite{Dam01}, the hidden subgroup problem
\cite{Kuperberg03},  and for solving linear systems of     
equations \cite{Harrow09}. A quantum walk can traverse a graph exponentially
faster than any classical random walk which allows for the efficient
solution of certain oracle problems  \cite{Childs02}.  Aharonov et
al.~proposed a quantum algorithm which efficiently approximates the Jones polynomial at
any primitive root of unity \cite{Aharonov06}. 

It seems to be clear that quantum entanglement and quantum
interference are two key resources which provide for the enhanced
information processing capabilities of quantum systems
\cite{Bennett00}.  But in spite 
of the many known examples in which quantum
information processing outperforms classical information processing, it is
not entirely clear how exactly  
these resources enable the speed of quantum algorithms, nor what the
largest possible speed-up is. 
It was shown \cite{Jozsa03} that a unitary quantum
algorithm in which 
entanglement remains ``$p$-blocked'' (i.e.~the number of qubits which
at any time are entangled is not larger than $p$), can be efficiently
simulated classically.
Nevertheless, the same authors argued that it might be misleading to
consider entanglement as the key resource.  As long as the mechanism is not    
identified by which any specific quantity creates the
speed--up, one might suspect its creation in large amounts rather correlated
with the quantum 
acceleration than being its cause. 
Entanglement is definitely crucial for tasks like quantum teleportation
\cite{Bennett93}, where its role can be
understood through the enhanced correlations between subsystems that
quantum mechanics can provide. 

Recently, 
experimental implementations \cite{Bigourd08} of factoring integer numbers
using Gauss sums, have re-emphasized the role of interference in
quantum computation. While these methods do not appear to be scalable
to integers with many 100 digits, and can be implemented with
classical waves, they are reminiscent of simple quantum algorithms
like the Deutsch-Jozsa algorithm, in which interference is clearly
seen at work.
Contrary to quantum
entanglement, quantum interference has been surprisingly little
studied.  From a physicist's perspective, 
quantum interference is an effect that arises from the coherent
superposition of quantum mechanical wave functions.  This can lead to
interference maxima and minima in probability distributions, as is
well--known from quantum particles going through a double-slit,
electrons in a mesocscopic solid state
circuit
\cite{Sharvin89}, or
interfering Bose-Einstein condensates 
\cite{Andrews97}. Quantum interference can also focus  
the probability distribution in a computer over its possible states at
the outcome of a calculation onto the state corresponding to the
result of the calculation. Without the coherence of quantum superpositions,
probabilities can 
only be propagated classically, i.e.~through a stochastic map, which
is, of course, void of any interference effects. If we want to
quantify interference, we therefore have to quantify to what extent
the propagation is coherent, as otherwise there is no telling if the
production of a final probability distribution involved interference
or not. But coherent propagation alone is not tantamount to
interference. At least two wave functions have to be
superposed in order to create interference. Very generally, one would
want to 
attribute more interference to a process in which many waves get
superposed with similar weights than to one where only very few waves
contribute. This implies a basis
dependence of interference, as a superposition in one basis is a
single basis state in another.  

In \cite{Braun06} a measure of quantum interference was introduced which allows to quantify interference in any quantum
mechanical process in a finite dimensional Hilbert space. Any such
process can be described by a completely positive map $P$ that maps an
initial density matrix $\rho$ to a final one, 
$\rho'=P\rho$. Written in a given basis, where $\rho$ and $\rho'$ have matrix
elements $\rho_{mn}$ and $\rho'_{kl}$, respectively, we have
$\rho'_{mn}=\sum_{kl}P_{mnkl}\rho_{kl}$. In that basis, the
interference associated with the positive map $P$ is written as
\begin{equation} \label{IM}
\cI(P)=\sum_{i,k,l}|P_{ii,kl}|^2-\sum_{i,k}|P_{ii,kk}|^2\,.
\end{equation}
While this interference measure may not be unique, it has the desired
property of measuring the coherence and the ``equipartition'' of
superposed basis states. Indeed, if $P$ reduces to a classical
stochastic map, it only propagates initial probabilities
$\rho_{ii}$ to final ones, $\rho_{kk}$.  Exactly the terms
responsible for this classical process are subtracted out in
eq.(\ref{IM}), such that if no coherences are propagated to final
probabilities, we have zero interference. The squares of the matrix
elements of $P$ in (\ref{IM}) allow to measure the equipartition
property, as is seen most easily for purely unitary propagation, where
$I$ reduces to $N-\sum_{ij}|U_{ij}|^4$, where $U$ is the unitary matrix
propagating the wave function, and $N$ the dimension of Hilbert
space. Perfectly equipartitioned unitary matrices
($|U_{ij}|=1/\sqrt{N}$) create the maximum amount of interference
possible for unitary propagation, $\cI=N-1$. As an example, the Hadamard gate creates one bit of interference, an
``i-bit''. Both Shor's and Grover's algorithm create an exponential amount of
interference (in the number of qubits). The part of the quantum
algorithm after application of the initial Hadamard gates creates only
about three i-bits in Grover's algorithm, but still an exponentially
large amount of interference in Shor's algorithm. If the success
probability of these algorithms is lowered by introducing unitary
errors or decoherence, so is in general the
interference \cite{BraunG08}. For unitary quantum algorithms randomly
drawn from the Circular Unitary Ensemble (CUE), interference is very
narrowly distributed about the mean value, which itself is almost the
maximum possible value \cite{Arnaud07}.  In other words, almost all
unitary quantum 
algorithms lead to an exponentially large amount of interference. This
situation is 
reminiscent of entanglement, as almost all states of high-dimensional
bipartite systems are close to maximally entangled \cite{Hayden06}.  

It also turned also,
however, that quantum interference is {\em not} necessary for several
tasks. For example the transmission of a quantum state through a chain
of qubits needs only a very small amount of interference
\cite{Lyakhov07}.  And cloning of a quantum state can be performed
just as well without interference as with interference
\cite{Roubert08}.  

Almost all investigations of interference have focused so far on
unitary propagation.  Recently the benefits of more general, partly
dissipative and decoherent evolutions have been emphasized, both in
the context of quantum enhanced measurements \cite{Braun09}, as in
quantum computing \cite{Verstraete09}. Moving on in this direction, we
investigate in this paper the statistical properties of interference
for general positive maps.  We construct such maps by propagating
unitarily a central system and an environment, which we take here both
as finite dimensional quantum systems, and then tracing out the
environment. We calculate analytically the first and second moments of
the distribution. We first focus on an environment that consists of a
single spin (such as a an ancillary qubit or qudit), and 
generalize then to an arbitrary number of spins, all taken initially
in a thermal state at arbitrary temperature.  We also calculate
numerically the entire interference distribution for small system
sizes. 

\section{Statistics of interference for a quantum system coupled to a
  single spin}
In this section we first review the propagation of a
finite dimensional quantum system that interacts with an arbitrary environment
consisting of another finite dimensional quantum system. The corresponding
propagator is a completely positive map of the
initial density matrix of the system to its final density matrix
\cite{Nielsen00}. While a
finite dimensional environment does not constitute a true
heat--bath in the sense of inducing irreversible behavior,  the study
of such a simple situation is motivated by quantum information theory,
where one frequently encounters ancilla qubits that are added to the
main quantum information processor.  Furthermore, the tracing out of any
environment with dimension larger than one does lead to decoherence as
soon as the system and its environment become correlated or
entangled, such that we will be able to study quantitatively the
influence of decoherence on quantum interference.  Further freedom
lies in the choice of the initial state of the environment, which can
be in a mixed state, e.g.~a thermal state reached by interaction with
its own heat-bath. We then derive the expression for the interference
of a quantum system whose environment is a simple spin initially in
thermal equilibrium, and study the statistical properties of the
interference of the completely positive map of the system under joint
unitary evolution of system and environment.  

\subsection{Propagator for a completely positive map }

Consider a bipartite system consisting of a
system ${\cal S}$ (Hilbert space ${\cal H}_S$ with dimension $n$) and
an environment 
${\cal E}$ (Hilbert space ${\cal H}_E$ with dimension $m$). Let $W$
and $W'$ be the initial and final density matrices
of the total system, respectively. We consider an initial product state
$W=\sigma\otimes\epsilon$ of the density matrices $\sigma$ and $\epsilon$ of the
system and its environment, respectively.  Under the condition that the
total system ``${\cal S}+{\cal E}$'' can be considered 
closed on the time scale of the evolution we are interested in, the
evolution of the system and its environment in the tensor 
product Hilbert space ${\cal H}_S\otimes{\cal H}_E$ of dimension
$N=n\times m$ is purely 
unitary and can be
represented by a unitary matrix $U$, $W'=UWU^{\dagger}$. 
In components we have  
\begin{eqnarray*}
W'_{\alpha_1\alpha_2,\beta_1\beta_2}&=&\sum_{\gamma_1,
  \gamma_2,\delta_1,
  \delta_2}^{n,m}U_{\alpha_1\alpha_2,\gamma_1\gamma_2}W_{\gamma_1\gamma_2,
  \delta_1\delta_2}U^{*}_{\beta_1\beta_2,\delta_1\delta_2} \label{W'} 
\end{eqnarray*}
where the indices with subscripts 1 and 2 label the basis
states of the system and the environment, respectively. The final
reduced density matrix of the system is found by tracing out the
environment, $\rho'=\tr_{\cal E}W'$, or, explicitly,
$\rho'_{\alpha_1\beta_1}=
\sum_{\alpha_2}^{m}W'_{\alpha_1\alpha2,\beta_1\alpha_2}\,.
$ 
From (\ref{W'}) and the initial 
$W_{\gamma_1\gamma_2}=\rho_{\gamma_1\delta_1}\epsilon_{\gamma_2\delta_2}$
we obtain  the propagation of ${\cal S}$ alone,
\begin{eqnarray*}
\rho'_{\alpha_1\beta_1}&=&\sum_{\gamma_1,
  \delta_1}^{n}P_{\alpha_1\beta_1,\gamma_1\delta_1}\,\rho_{\gamma_1\delta_1}\label{rho'} 
\end{eqnarray*} 
where the components of the propagator are given by
\be
P_{\alpha \beta, \gamma \delta}=\sum_{\mu,\nu,\rho}^{m}U_{\alpha \mu,\gamma \nu}\epsilon_{\nu \rho}U_{\beta \mu,\delta \rho}^{*}
\ee
This propagator $P$ is a superoperator that maps the initial density
operator $\rho$ to the final
density operator $\rho'$.  The procedure of "hamiltonian embedding" we
have used guarantees that this propagator is a completely positive map
\cite{Nielsen00}. As expected, $P$ depends
not only on $U$ but also on the 
initial state of the environment $\epsilon$. We are now in a position
to calculate the interference for the propagation (\ref{rho'}). To
obtain explicit results, we consider particular initial states
for the environment.  We 
start with a single spin in thermal equilibrium, and later generalize
to several spins in thermal equilibrium.

\subsection{Interference in a quantum system coupled to a single spin
  in thermal equilibrium } 

Consider the situation where the environment is a single spin of size
$(d-1)/2$, which corresponds to a Hilbert space of dimension $m=d$. We 
assume that the energy 
levels of the spins are equally spaced, with neighboring levels
separated by an energy 
$\hbar\Omega$, as is 
the case for atomic or nuclear spins under linear Zeeman effect
in an external magnetic field. In its own eigenbasis, the matrix
elements of the spin
Hamiltonian $H^{(1)}$ reads
\be
H_{\nu \rho}^{(1)}=\hbar\Omega\,\nu\,\delta_{\nu \rho}\label{h1}
\ee
where $1\leq\nu\leq d$ and $\nu-1$ is the number of excitations of the
spin. We choose the spin to be initially at thermal equilibrium at
temperature $T=\frac{1}{k_{B}\beta}$, such that its density matrix can
be written as 
\be
\epsilon=\frac{e^{-\beta H}}{\tr(e^{-\beta H})}\to\epsilon_{\nu \rho}=\frac{1}{Z}e^{-\beta\,\hbar\,\Omega \nu}\delta_{\nu \rho}
\ee
with partition function 
\be 
Z\equiv Z(x)=\sum_{\nu}^{d}e^{-\beta\,\hbar\,\Omega \nu} 
=\frac{1-e^{-d \,x}}{e^{x}-1}\,,
\ee
and $x=\beta\,\hbar\,\Omega$. The propagator $P$ simplifies,
\be
P_{\alpha \beta, \gamma \delta}=\frac{1}{Z}\sum_{\mu,\nu}^{d}U_{\alpha \mu,\gamma \nu}U_{\beta \mu,\delta \nu}^{*}e^{-x\,\nu}\,.\label{Pfin}
\ee
Inserting (\ref{Pfin}) into (\ref{IM}), we finally obtain the
expression for the interference in the propagation of ${\cal S}$ alone,
\begin{eqnarray}\label{inter}
\I&=&\sum_{\alpha,\gamma\neq\delta}^{n}|P_{\alpha \alpha, \gamma
  \delta}|^2=\frac{1}{Z^2}\sum_{\alpha,\gamma\neq\delta}^{n}|\sum_{\mu,\nu}^{d}U_{\alpha \mu,\gamma \nu}U_{\alpha\mu,\delta\nu}^{*}e^{-x\,\nu}|^2 \nonumber \\ 
&=&\frac{1}{Z^2}\sum_{\alpha,\gamma\neq\delta}^{n}\sum_{\mu,\nu,\rho,\sigma}^{d}e^{-x\,(\nu+\sigma)}U_{\alpha \mu,\gamma \nu}U_{\alpha \mu,\delta \nu}^{*}U_{\alpha \rho,\gamma \sigma}^{*}U_{\alpha \rho,\delta \sigma}\,. \nonumber
\end{eqnarray}
We are now in the position to investigate the statistical properties
of $\cI$ based on the statistics of $U$.  Without prior knowledge of a
particular set of quantum algorithms or physical time evolution, it is
natural to choose $U$ uniformly distributed with respect to the 
Haar measure $dU$ of the unitary group $U(N)$. The statistical
ensemble for the joint propagator of system and environment is then
the well  
known "Circular Unitary Ensemble" (CUE). This allows us in particular
to recover previously known results \cite{Arnaud07} for the
interference statistics 
for unitary propagation of ${\cal S}$ in the limit where the dimension
of the environment is reduced to one, as we will show below. 

\subsection{Numerical results}\label{sec.numerical}

\begin{figure}[h]
\epsfig{file=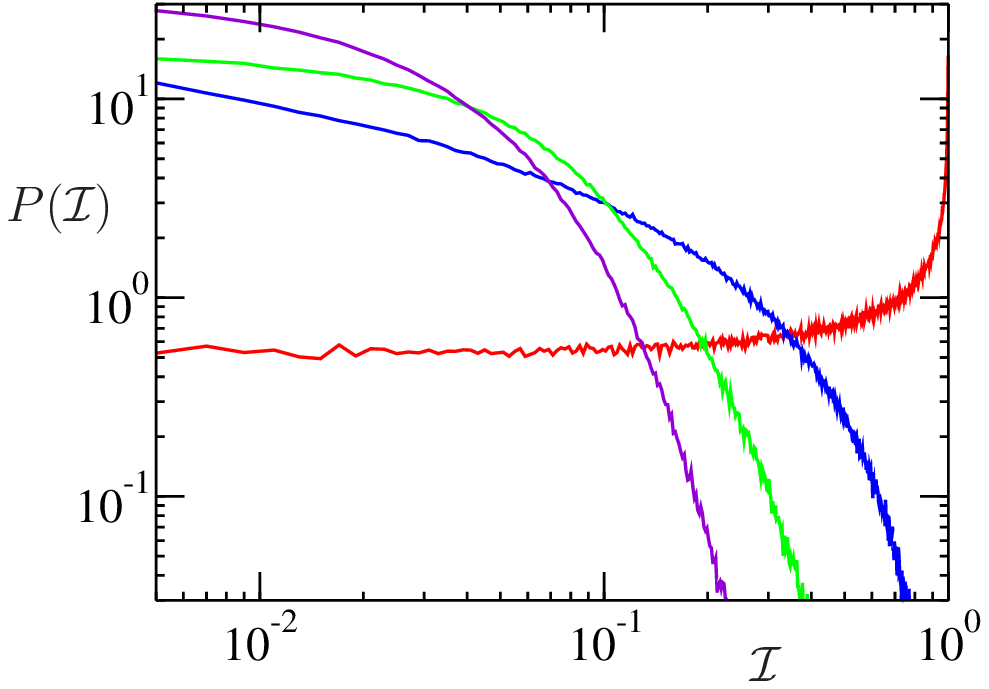,width=5cm,angle=0}
\epsfig{file=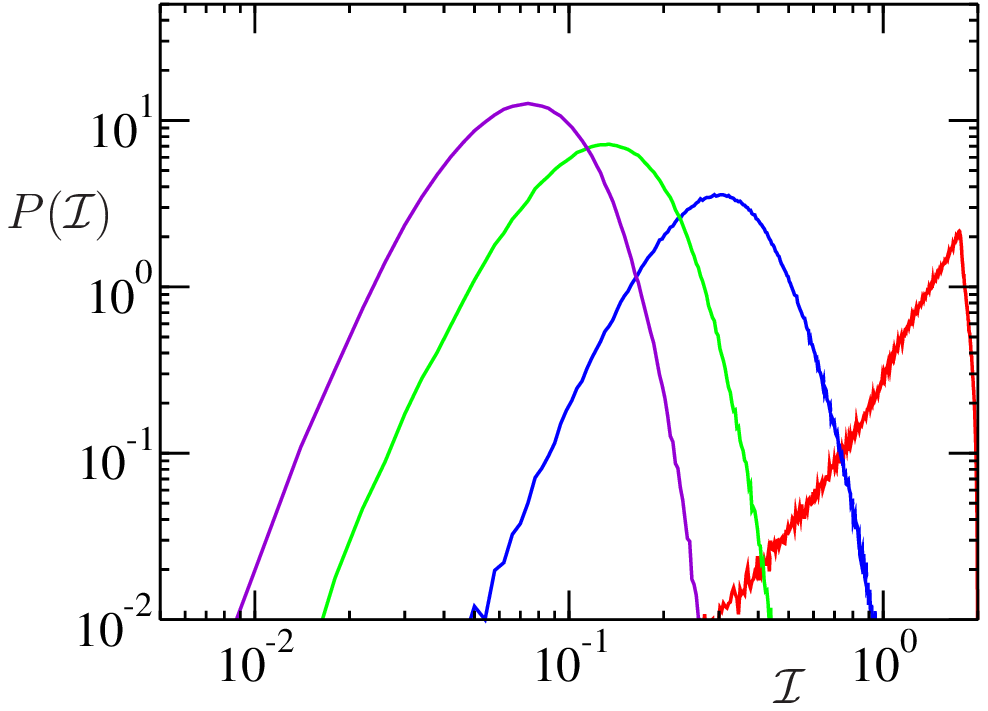,width=5cm,angle=0}
\epsfig{file=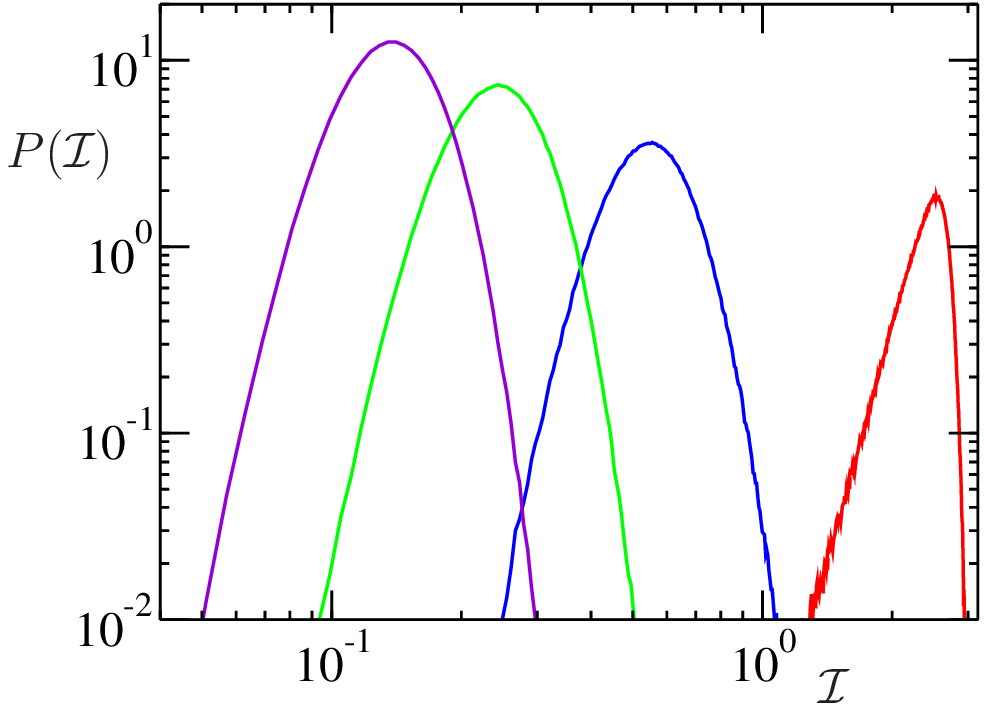,width=5cm,angle=0}\\
\caption{(Color online) Numerically calculated interference distributions on
  a log-log scale for $x=0.1$ for $n=2,3,4$ (from the left to the right). In
  each plot, $m=1,2,3,4$ from the right to the left (red, blue, green,
  purple, respectively). The number of realizations is $n_r=10^6$ in all
  cases.\\} 
\label{fig.num} 
\end{figure}

For small dimensions $n$ and $m$, one can obtain the entire distribution of
interference $P(\cI)$ numerically. We have produced numerically
unitary matrices of size $N=n\times m$ drawn from CUE using Hurwitz
parametrization \cite{Hurwitz1897,Pozniak98}. In order to obtain good statistics
we have used 
$10^6$ matrices for the calculation of the distribution. Figure
\ref{fig.num} shows $P(\cI)$ for systems with sizes from $n=2$ to 4, coupled to
an environment of size $m=1$ to 4 at inverse temperature $x=0.1$.
\begin{figure}[h]
\epsfig{file=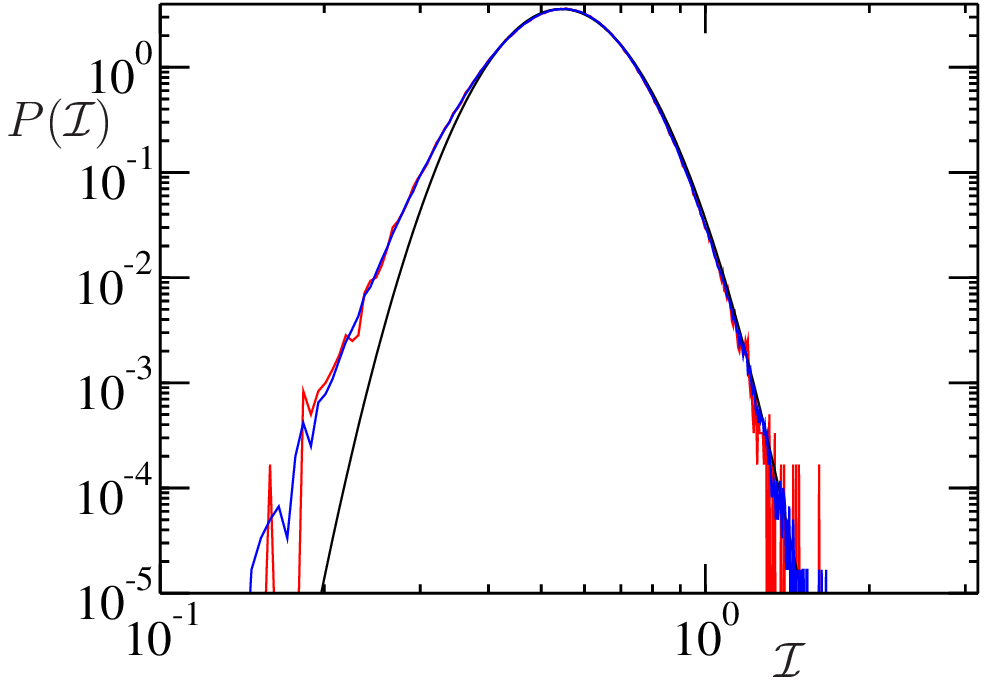,width=5cm,angle=0}
\epsfig{file=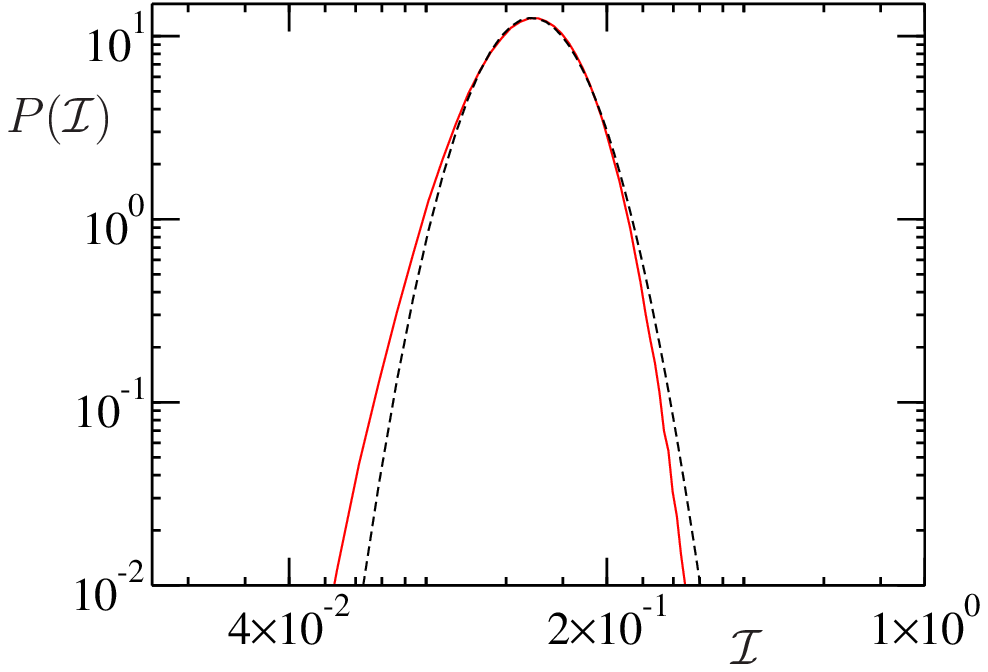,width=5cm,angle=0}
\epsfig{file=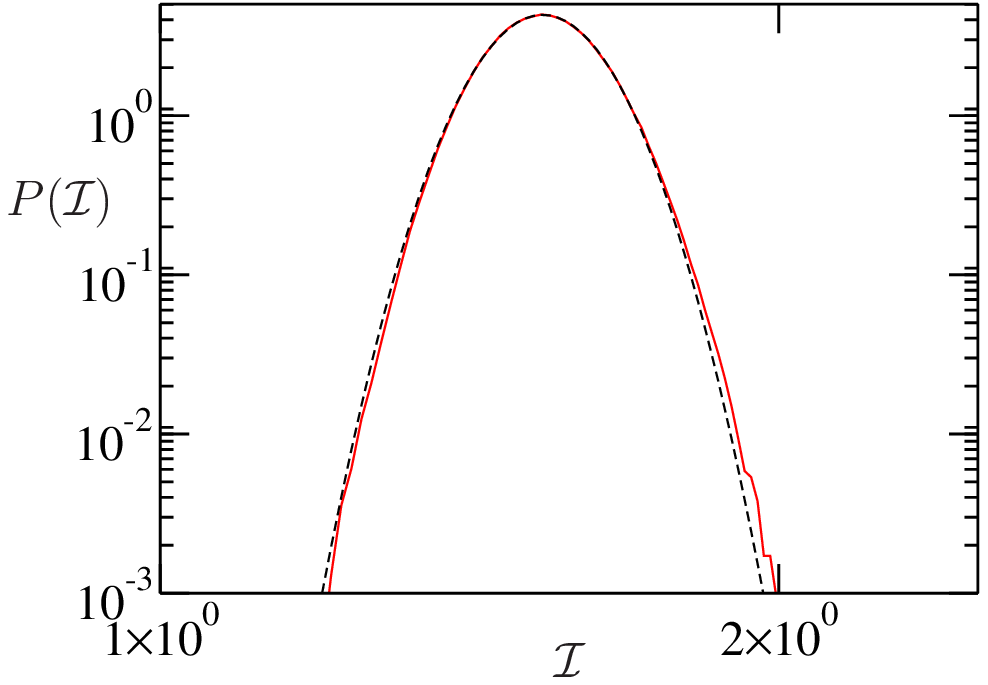,width=5cm,angle=0}
\caption{(Color online) Fit of numerically calculated $P(\cI)$ (red curves)
  to a log-normal distribution (black dashed curves)  at x = 0.1 for
  (n,m)=(4,2), (4,4) and (8,2) (from left to right). All fits are for
  $n_r=10^6$ except for the first plot in which the blue curve is for
  $n_r=10^7$.  } 
\label{fig.num2} 
\end{figure} 
In the case $n=2$, where the analytical calculation is possible the
distributions are very wide 
(see \cite{Arnaud07}  for $m=1$ where $P(\I)=\frac{1}{2\sqrt{1-\I}}$). For
higher  
values of $n$, the distribution becomes more and more peaked, and,  on a
log-log scale, more and 
more symmetric with respect to the maximum.   
The tails of the distribution decay more rapidly in the non-unitary case.
We see that both the most probable value of $\cI$ and the width of
the peak decrease 
when $m$ increases. As expected, the decoherence due to the coupling to
the environment destroys the interference more efficiently with increasing
$m$. For fixed $m$ the general distribution behaves qualitatively like the
distribution for the unitary case studied in \cite{Arnaud07}, i.e.~the most
probable interference and mean interference increase with $n$, 
whereas the width of the distribution decreases with increasing $n$. 
A change of the temperature essentially shifts the
distribution. This is due to a change of the average interference with the
temperature as we will see later (see eq.(\ref{Imoyfinal})), and justifies
why we have plotted all distributions for $x=0.1$. 

Fig.\ref{fig.num2} shows that $P(\cI)$ is well fitted by a log-normal
distribution, 
\begin{equation} 
P(\cI)=\frac{\exp\left(\frac{(\log(\cI)-\mu)^2}{2\sigma^2}\right)}{\cI\sqrt{2\pi}\sigma}\,. 
\end{equation}
The fits work particularly well close to the center of the distributions,
whereas deviations appear in the wings of the distribution. In addition, the
wings appear to be clipped, but this is at least partly an effect of the
finite number of realizations available.  This is visible from the example
$n=4, m=2$, where we have increased the number of realizations $n_r$ from
$10^6$ to  $10^7$.  In the latter case, the clipping appears at
substantially larger values of $\cI$. 
 
The numerically obtained distributions suggest that $P(\I)$ is for
$n>2$ well characterized by its first and second moments. 
We will now present
analytical results for these two moments which confirm the qualitative
observations above for arbitrary values of $n$, $m$, and $x$, and make them
more quantitative. 

\subsection{Analytical results}

\subsubsection{Average interference}
The average interference $\langle \cI\rangle$ follows from eq.(\ref{inter}),
\begin{eqnarray}\label{Imoy1}
\bra \I
\ket&=&\frac{1}{Z^2}\sum_{\alpha,\gamma\neq\delta}^{n}\sum_{\mu,\nu,\rho,\sigma}^{d}e^{-x\,(\nu+\sigma)}\bra
U_{\alpha \mu,\gamma \nu}U_{\alpha \mu,\delta \nu}^{*}U_{\alpha
  \rho,\gamma \sigma}^{*}U_{\alpha \rho,\delta \sigma}\ket 
\end{eqnarray}
where $\langle\,.\rangle\equiv \int dU (.)$ means average over 
CUE. For the monomials composed of a relatively small number of factors
$U_{\alpha \mu,\gamma 
  \nu}$ to be averaged here, the technique of invariant integration is
well suited. 
We use the diagrammatical language introduced in
\cite{Aubert03,Aubert04} to express $\langle \cI\rangle$ as
\begin{eqnarray}
\cI&=&\frac{1}{Z^2}\sum_{\alpha,\gamma\neq\delta}^{n}\sum_{\mu,\nu,\rho,\sigma}^{d}e^{-x\,(\nu+\sigma)}\Fmax{$\alpha
  \mu$}{$\alpha \rho$}{$\gamma \nu$}{$\delta \nu$}{$\gamma
  \sigma$}{$\delta \sigma$}\label{Iavdiag} 
\end{eqnarray}
We refer the reader to \cite{Aubert03,Aubert04,Braun06c} for a detailed
explanation and derivation of this technique, but summarize here the
main features. For the sake of clarity we revert momentarily to single
roman indices $i$, $j$ 
etc.~for rows and columns. 
All distinct row (column) indices that appear in the matrix elements of
the monomial are 
represented by vertices on the left (right) with the corresponding label,
irrespectively of whether or not they arise from a matrix element
$U_{ij}$ or its complex conjugate $U_{ij}^{*}$. A complex conjugate
factor $U_{ij}^{*}$ is then represented by a thin
solid line between the vertices $i$ and $j$, whereas a factor $U_{kl}$ is
represented by a dotted line between the vertices $k$ and $l$. When a
given matrix element occurs with multiplicity $t$, a single line
is drawn with the 
number $t$ next to it to keep track of the multiplicity. Factors like
$|U_{ij}|^2$ are
represented by thick solid lines, which can also have a multiplicity
larger than one.
In \cite{Aubert03} it was shown that the invariance of the
Haar measure under arbitrary unitary transformations leads to the 
following important properties:  
\begin{enumerate}
\item[(a)] The value of a diagram does not depend on the specific
  values of the vertices. It only depends on the form of the
  diagram. This means that diagrams can be drawn without specifying
  the explicit 
  values of the vertices. For example,
\begin{equation}
\bra U_{11}U_{11}^{*}U_{12}U_{12}^{*} \ket=\bra
U_{24}U_{24}^{*}U_{26}U_{26}^{*} \ket=\FFuu \,.
\end{equation}
\item[(b)] If for at least one vertex in the diagram, the number of
  thin solid lines that originates from the vertex differs from the number
  of dotted lines then the value of the diagram is zero. For example,
\begin{eqnarray}
\bra U_{11}U_{12}^{*}U_{23}^{*}U_{24} \ket&=&\FFmax=0\\
\bra U_{11}U_{12}^{*}U_{21}U_{22}^{*} \ket&=&\EEd\neq 0 \,.
\end{eqnarray}
\end{enumerate}
In eq.(\ref{Iavdiag}), we sum over all row and column indices and different
type of diagrams therefore appear, depending on which vertices coincide.
Combinations of indices contribute for which the
vertices  
($\gamma\nu$) and ($\delta\nu$) collapse on the vertices
($\gamma\sigma$) and ($\delta\sigma$), respectively,
i.e.~configurations with
$\nu=\sigma$. We thus have 
\begin{eqnarray}
\bra \I \ket&=&\frac{1}{Z^2}\sum_{\alpha,\gamma\neq\delta}^{n}\sum_{\mu,\nu,\rho}^{d}e^{-2 x\,\nu}\Ed{$\alpha \mu$}{$\alpha \rho$}{$\gamma \nu$}{$\delta \nu$}\\
&=&\frac{1}{Z^2}\sum_{\alpha,\gamma\neq\delta}^{n}\Big(\sum_{\nu}^{d}e^{-2
  x\,\nu}\Big)\Big(\sum_{\mu=\rho}^{d}\Fuu{$\alpha \mu$}{$\gamma
  \nu$}{$\delta \nu$}+\sum_{\mu\neq\rho}^{d}\Ed{$\alpha \mu$}{$\alpha
  \rho$}{$\gamma \nu$}{$\delta \nu$}\Big)\,. 
\end{eqnarray}
At this point only two types of diagrams remain, and since their values
do not depend on the summation indices, we get
\begin{eqnarray}
\bra \I \ket&=&\frac{Z(2x)}{Z^2(x)}\, n^{2}(n-1)\Big(d\FFuu+d(d-1)\EEd\Big)\,.\label{Iavfin}
\end{eqnarray}
The values of the two diagrams are easily found \cite{Aubert03},
\begin{eqnarray*}
\FFuu&=&\frac{1}{N(N+1)}\\
\EEd&=&\frac{1}{N(N^2-1)}\,.
\end{eqnarray*}
The prefactor can be rewritten as
\begin{equation} \label{h}
h(x)\equiv\frac{Z(2x)}{Z^2(x)}=\coth(dx/2)\tanh(x/2),
\end{equation}
 and we finally
obtain, with $d=m$, 
\be \label{Imoyfinal}
\bra \I(n,m,x)
\ket=\coth(\frac{mx}{2})\tanh(\frac{x}{2})\frac{nm(n-1)^2}{(n^2m^2-1)}\,. 
\ee
This is our first central result which we now discuss in detail.

We first observe that the entire temperature dependence is entirely
contained in the prefactor $h(x)$.  Its limits for
$x\to 0$ and $x\to \infty$ are $1/m$ and $1$, respectively. In between,
$h(x)$ increases monotonously.  We thus find that the average
interference decreases with increasing temperature, an intuitively
appealing result. 
Only the dimension of the environment $m=d$ enters the dependence on
temperature. This is true in fact for all moments of $P(\cI)$, as the
entire temperature dependence is contained in 
factors $\exp(-x\nu)$ which are always summed over $\nu=1,\ldots,m$.     

In the particular case $m=1$, i.e.~$n=N$, we recover as expected
the expression for purely unitary propagation
\cite{Arnaud07}, 
\be
\bra \I(n,1,x)
\ket=\frac{N(N-1)^2}{N^2-1}
=\frac{N(N-1)}{N+1}=\bra
\I_{U}(N)\ket\,.
\ee
No entanglement or correlations with the environment can arise in this
case, as a single state always factors out, such that the dynamics of
${\cal S}$ remains indeed entirely unitary.  

\begin{figure}[h]
\epsfig{file=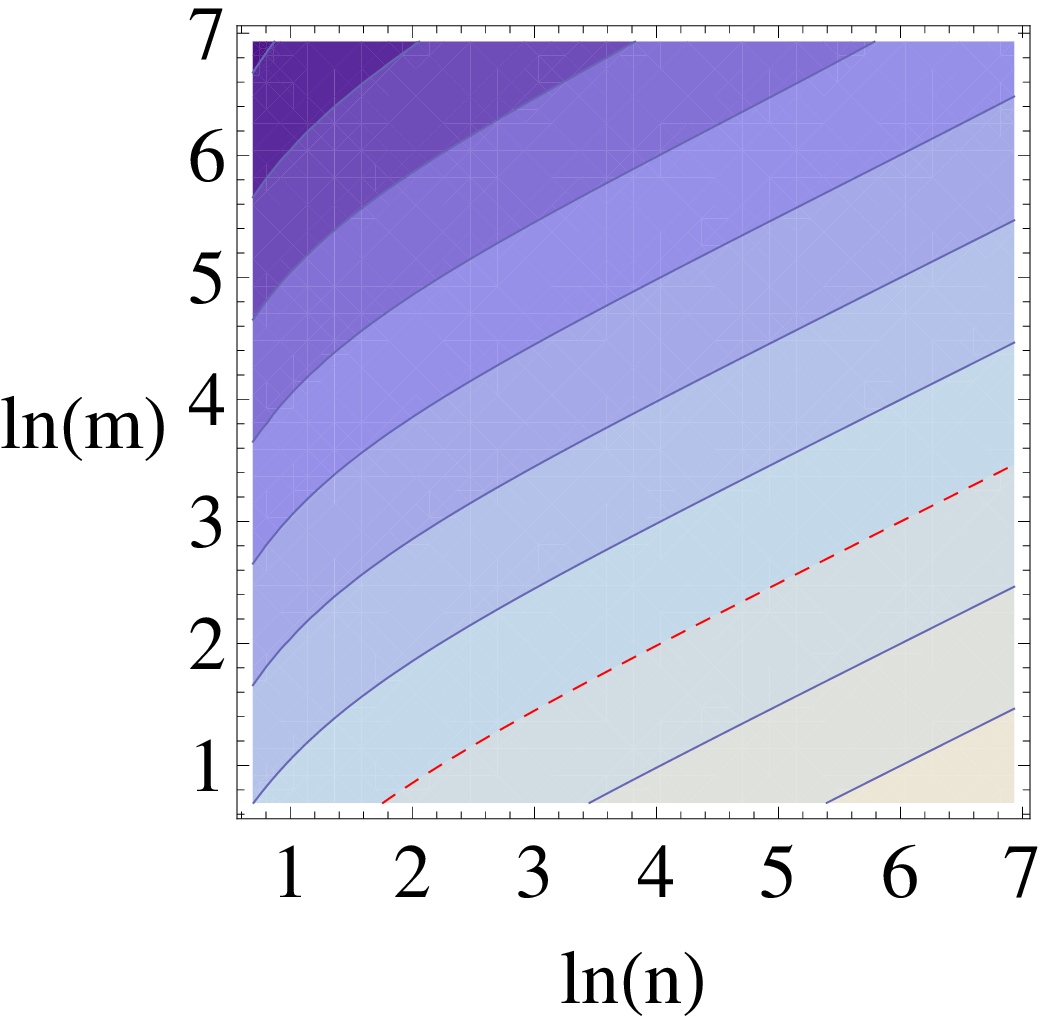,width=8cm,angle=0}
\epsfig{file=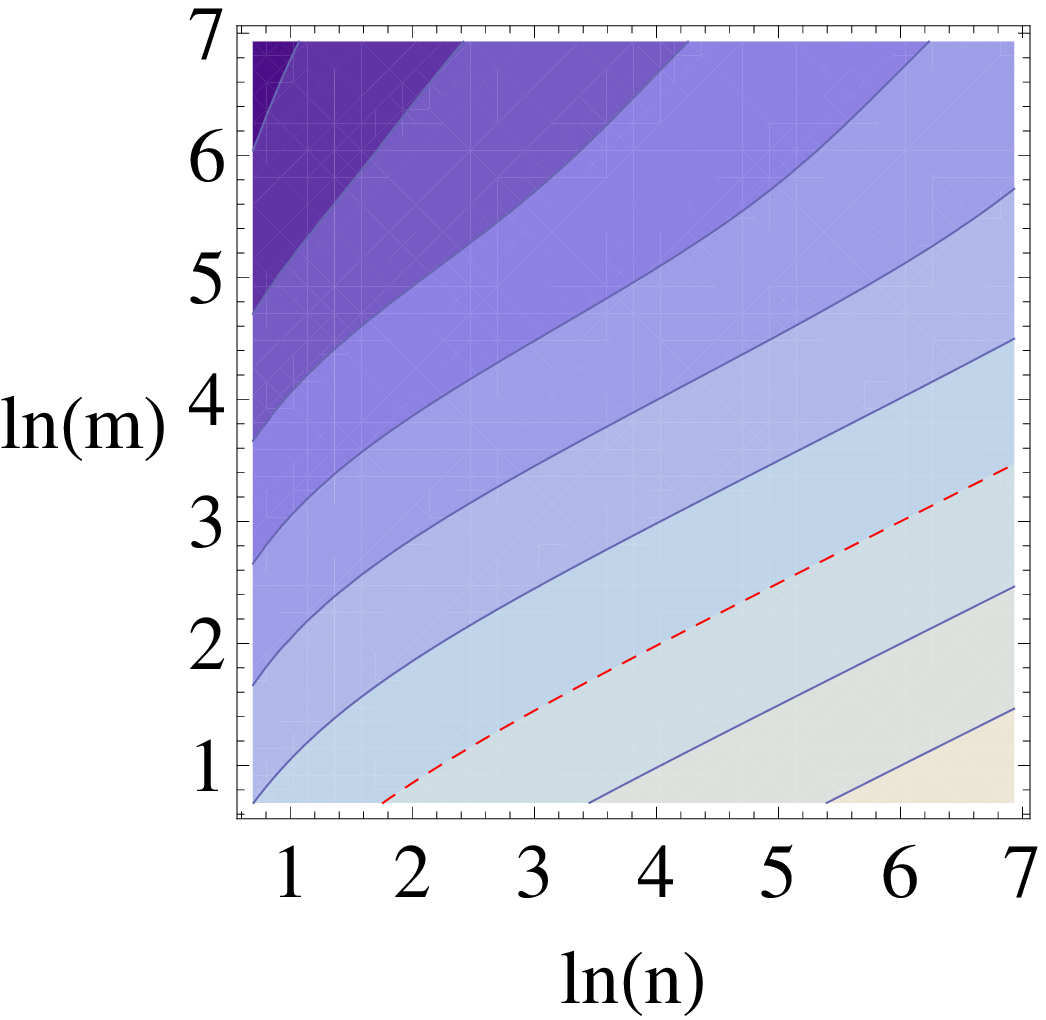,width=8cm,angle=0}\\
\epsfig{file=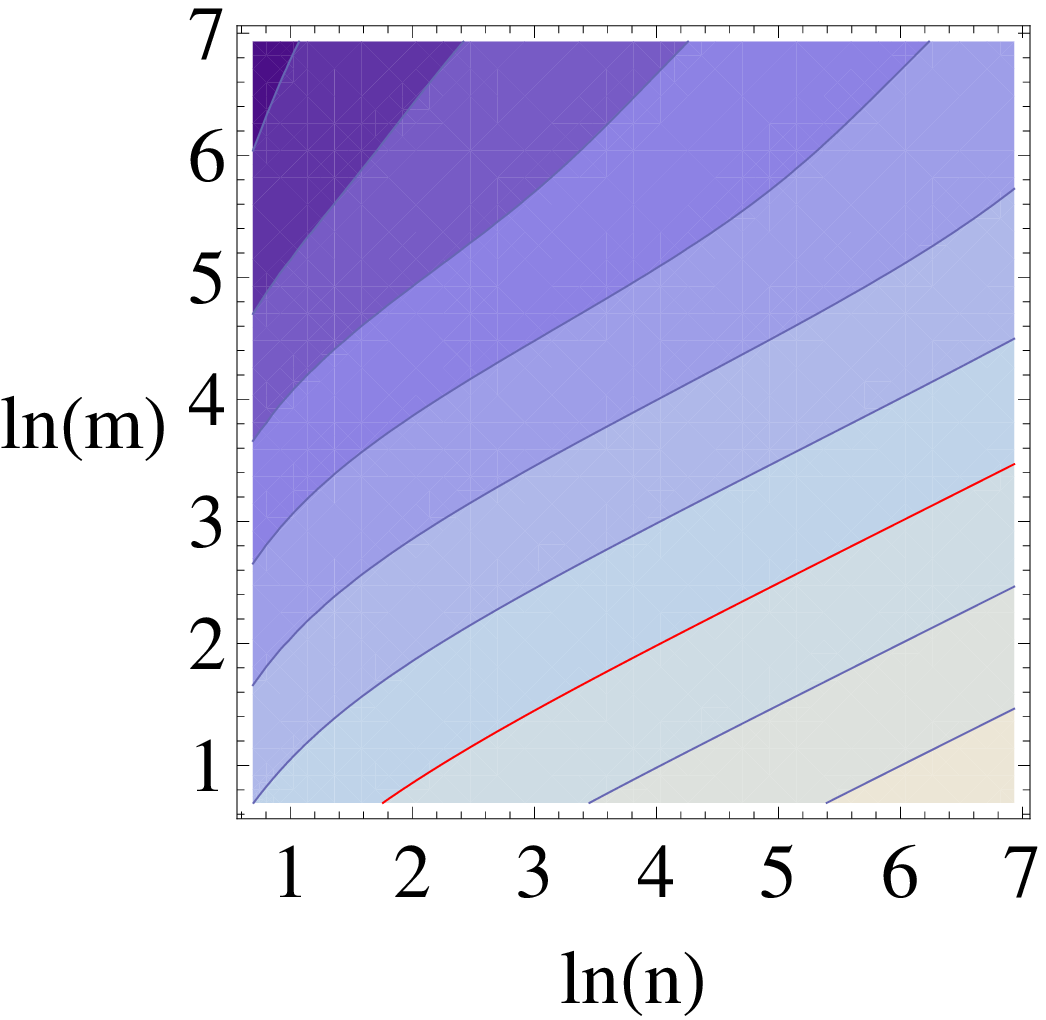,width=8cm,angle=0}
\epsfig{file=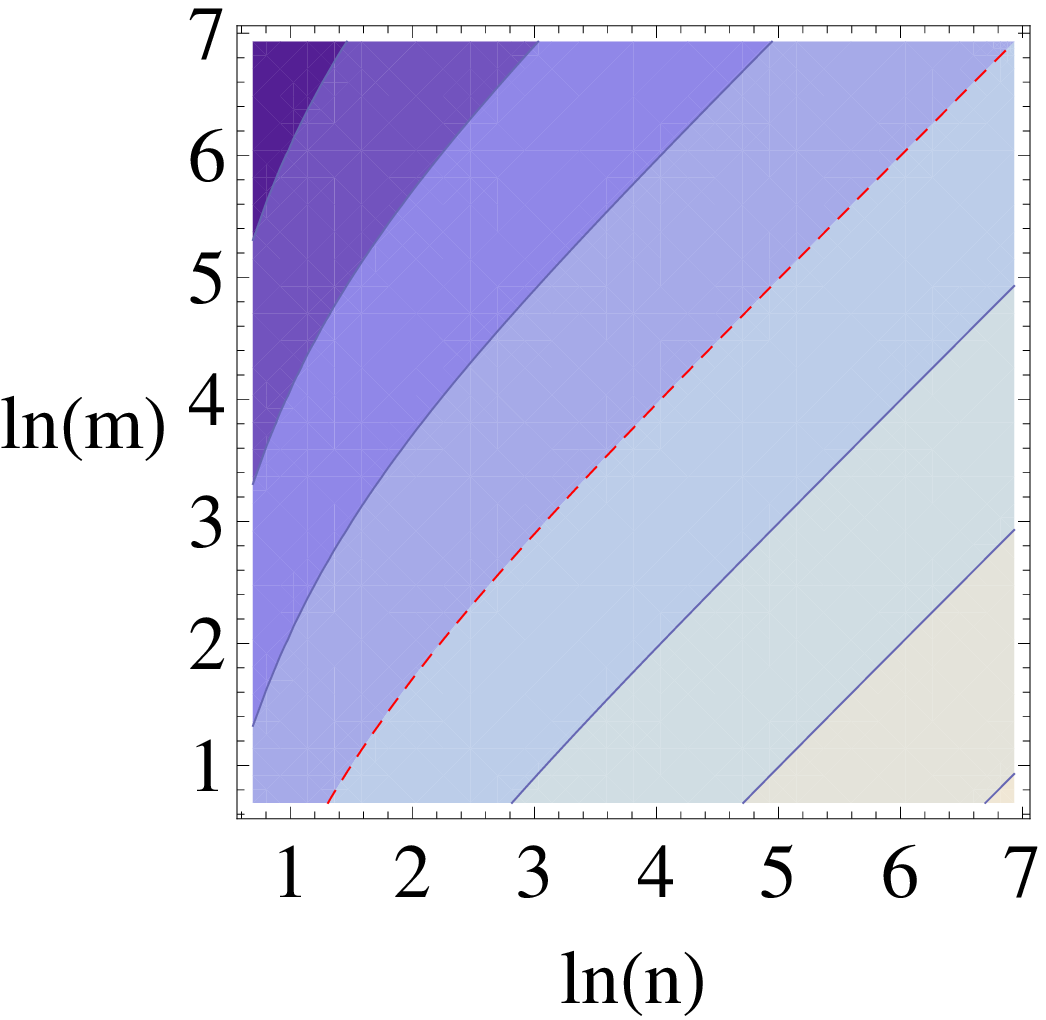,width=8cm,angle=0}
\caption{(Color online) Contour plot of $\ln(\bra \I(n,m) \ket)$ for x=0.001, 0.01, 0.1
  and 10 (upper left to lower right), for $n$ and $m$ between 2 and
  1024. The distance between the contours is 2, and the red dashed line is for
  $\ln(\bra \I(n,m) \ket)=0$. Values
  increase from  dark to bright colors.}
\label{fig.moy} 
\end{figure}

Contrary to what might be expected naively, the unitary result is
not recovered for zero 
temperature, $x\to \infty$.  Rather
one finds 
\begin{eqnarray}
\lim\limits_{x\to\infty}\bra \I(n,m,x)
\ket&=&\frac{N(n-1)^2}{N^2-1}\label{avIxInfty}\,, 
\end{eqnarray}
We recall that $N=n\times m$. 
For $n \gg 1$ and $m$ fixed we have the asymptotic behavior 
\begin{eqnarray}
\bra \I_{U}(n) \ket &=& n-2 + O(\frac{1}{n})\\
\bra \I(n,m,x\to\infty) \ket &=& \frac{n-2}{m} +
O(\frac{1}{n})\simeq\frac{\bra\I_{U}(n)\ket}{m} 
\end{eqnarray}
We see that for $n\gg 1$, the average interference still scales linearly with the system size, but is roughly a factor $m$ smaller than in the unitary case. The reason for this reduction is, of course, that even for a heat bath
initially in a pure ground state, the common unitary dynamics of ${\cal S}$ and
${\cal E}$ entangles ${\cal S}$ and ${\cal E}$, such that after tracing out the
environment non-unitary evolution of ${\cal S}$ results. The 
consequent loss of coherence manifests itself in a reduction of
interference.  
In the opposite limit of infinite temperature, $x\to 0$, the
temperature dependence of the prefactor $h(x)$ leads to 
reduction by another factor $m$, 
\begin{eqnarray}
\lim\limits_{x\to0}\bra \I(n,m,x)
\ket&=&\frac{n(n-1)^2}{N^2-1}\,.\label{avIx0}
\end{eqnarray}
The additional reduction is also seen in the asymptotic expansion for $n\gg
1$, which reads in this case  
\begin{eqnarray}
\bra \I(n,m,x=0) \ket &=& \frac{n-2}{m^2}
+O(\frac{1}{n})\simeq\frac{\bra\I_{U}(n)\ket}{m^2}\,.   
\end{eqnarray}
For $m \gg 1$ and $n$ fixed we find
\begin{eqnarray}\label{moyIxtoinf}
\bra \I(n,m,x\to\infty) \ket &=& \frac{(n-1)^2}{n m} + O(\frac{1}{m^3})\\\label{moyIxto0}
\bra \I(n,m,x=0) \ket &=& \frac{(n-1)^2}{n m^2} + O(\frac{1}{m^3})\,.
\end{eqnarray}
Eqs.(\ref{moyIxtoinf}) and (\ref{moyIxto0}) show that for fixed $n>1$, $\bra \I \ket$ decreases as $1/m$ ($1/m^2$) for zero temperature (infinite temperature).
In Fig.\ref{fig.moy} we plot $\bra \I(n,m,x)\ket$  for four different
  temperatures as function of $n$ and $m$. We see that for given temperature,
  $\bra \I(n,m,x)\ket$  increases with $n$, but decreases with
  $m$.  For large $n$, with $m$ and  $x$ fixed, the increase is
  essentially proportional to $n$, just as in the unitary case, albeit
  with a slope reduced by a factor $h(x)/m$.  For large $m$, with $n$ and $x$
  fixed such that $mx\gg 1$, the decrease of $\bra \I(n,m,x)\ket$ is
  roughly as $1/m$ 
  with a prefactor $\frac{(e^x-1)^2}{e^{2x}-1}\frac{n(n-1)^2}{n^2}$. 
More generally, an increase in the dimension of
the environment decreases the average interference in a power law
fashion with a power that crosses over from $m^{-2}$ for $x=0.001$ to $m^{-1}$ for x=10 and fixed $n$. One should not conclude from this, however, that a
quantum 
system coupled to an infinite dimensional heat bath will never show
any quantum interference effect.  Rather, it should be kept in mind
that we consider here generically strong couplings to the environment, in
the sense that a typical joint evolution operator $U$ of ${\cal S}$
and ${\cal E}$ does 
not distinguish the two subsystems, or, for that matter, a system
hamiltonian, bath hamiltonian, and coupling hamiltonian. It is natural
that such strong couplings destroy coherence and thus quantum
interference rapidly, but the situation can of course be different for
weak couplings.

\subsubsection{Second moment of the interference distribution}
In order to appreciate the width of the interference distribution as
function of $m,n$ and $x$, we now calculate the
second moment of $P(\cI)$.  
By taking the square of the eq.(\ref{inter}) we find 
\begin{eqnarray*}
\I^2
&=&\frac{1}{Z^4}\sum_{\alpha,\gamma\neq\delta}^{n}\sum_{a,g\neq
  d}^{n}\sum_{\mu,\rho,p,r}^{d}\sum_{\nu,\sigma,
  q,s}^{d}e^{-x\,(\nu+\sigma+n+s)}U_{\alpha \mu,\gamma \nu}U_{\alpha
  \mu,\delta \nu}^{*}U_{\alpha \rho,\gamma \sigma}^{*}U_{\alpha
  \rho,\delta \sigma}U_{ap,gq}^{*}U_{ap,dq}U_{ar,gs}U_{ar,ds}^{*}\\ 
\end{eqnarray*}
The fact that 8 factors $U$ appear now, makes the analytical
calculation of $\bra \I^2\ket$ rather cumbersome. As we will see,
altogether 19 different diagrams contribute.  We give here a rough
outline of the derivation, relegating most details and in particular
the values of all diagrams to the Appendix. In order to streamline the
presentation we introduce the following simplifications of notation:
\begin{itemize}
\item First, for the six subsystem indices, we substitute $(\alpha,\gamma,\delta,a,g,d)\to(\alpha_{1},\alpha_{2},\alpha_{3},\alpha_{4},\alpha_{5},\alpha_{6})$.
\item Similarly, for the eight environment indices, we replace
  $(\mu,\rho,p,r,\nu,\sigma,q,s)\to(\mu_{1},\mu_{2},\mu_{3},\mu_{4},\mu_{5},\mu_{6},\mu_{7},\mu_{8})$. 
\item We then drop the redundant letters $\alpha$ and $\mu$
  altogether, both from matrix elements and the diagrams. I.e.~we
  write matrix elements $U_{\alpha_i \mu_j,\alpha_k 
  \mu_l}$ just as $U_{ij,kl}$. So now $U_{11,11}$ 
is not the first 
  element of the matrix but it is the element with indices
  ($\alpha_{1}\mu_{1},\alpha_{1}\mu_{1}$). Recall that all $\alpha$
  ($\mu$) indices take values 
  between $1$ and $n$ ($m$), respectively. 
\item The constraints $\gamma\ne \delta$ and $g\ne d$ read now
  $\alpha_2 \neq \alpha_3$  and $\alpha_5 \neq \alpha_6$. They are assumed
  implicitly. 
\item We also make it a rule that in a sum $\{\al_i,\mu_j\}$ denotes
  the set of all 
  indices which appear explicitly in the summand as indices of matrix
  elements, or, equivalently, as labels of vertices, with 
  the exception of   those which appear under another sum 
   in the same expression. E.g.~in
  $\sum_{\{\alpha_i,\mu_j\}}\sum_{\mu_5,\mu_7}$, the first sum is over all
  $\alpha$'s and all $\mu$'s that show up in the diagram summed over,
  with the exception of $\mu_5$ and $\mu_7$, which are considered
  separately.    
\end{itemize}
We can then write
\begin{eqnarray}
\bra \I^2 \ket&=&\frac{1}{Z^4}\sum_{\{\al_i,\mu_j\}}^{(n,m)}e^{-x\,(\mu_5+\mu_6+\mu_7+\mu_8)} \bra U_{11,35}U_{11,45}^{*}U_{12,36}^{*}U_{12,46}U_{23,57}^{*}U_{23,67}U_{24,58}U_{24,68}^{*}\ket\label{brai2}\\
&=&\frac{1}{Z^4}\sum_{\{\al_i,\mu_j\}}^{(n,d)}e^{-x\,(\mu_5+\mu_6+\mu_7+\mu_8)} \Dmax\,.
\end{eqnarray}
We re-emphasize that the indices of $U$ which appear in eq.(\ref{brai2}) are
indices of indices, e.g.~$U_{11,35}\equiv U_{\alpha_1\mu_1,\alpha_3\mu_5}$.
As for equation (\ref{Imoy1}), the only non--vanishing contributions
arise from diagrams without open ends. They correspond to three distinct
configurations of the summation indices, namely
$\mu_5=\mu_6 \textrm{ and } \mu_7=\mu_8$, or
$\alpha_3=\alpha_5,\alpha_4=\alpha_6,\mu_5=\mu_7 \textrm{, and }
  \mu_6=\mu_8$, or
$\alpha_3=\alpha_6,\alpha_4=\alpha_5,\mu_5=\mu_8 \textrm{, and }
  \mu_6=\mu_7$.
These three configurations give rise to three sums,
\begin{eqnarray}
\bra \I^2
\ket&=&\frac{1}{Z^4}\Big(\sum_{\{\al_i,\mu_j\}}^{(n,m)}\sum_{\mu_5,\mu_7}^{m}
e^{-2x\,(\mu_5+\mu_7)} 
\Dqq{1,1}{1,2}{2,3}{2,4}{3,5}{4,5}{5,7}{6,7}+\sum_{\{\al_i,\mu_j\}}^{(n,m)}\sum_{\alpha_3,\alpha_4}^{n}
\sum_{\mu_5\neq\mu_6}^{m}e^{-2x\,(\mu_5+\mu_6)}
\Dqq{1,1}{2,3}{1,2}{2,4}{3,5}{4,5}{3,6}{4,6}\nonumber\\ 
&
&+\sum_{\{\al_i,\mu_j\}}^{(n,m)}\sum_{\alpha_3,\alpha_4}^{n}\sum_{\mu_5\neq\mu_6}^{m}e^{-2x\,(\mu_5+\mu_6)}
\Dqq{1,1}{2,4}{1,2}{2,3}{3,5}{4,5}{3,6}{4,6}\Big)\,.
\end{eqnarray}
The last two terms are equal as can be seen by exchanging the
summation indices $\alpha_{3}\leftrightarrow \alpha_{4}$.  We are
therefore left with
\begin{eqnarray} 
\bra \I^2\ket
&=&\frac{1}{Z^4}\Big(\sum_{\{\al_i,\mu_j\}}^{(n,m)}\sum_{\mu_5\mu_7}^{m}e^{-2x\,(\mu_5+\mu_7)}
\Dqq{1,1}{1,2}{2,3}{2,4}{3,5}{4,5}{5,7}{6,7}+2\sum_{\{\al_i,\mu_j\}}^{(n,m)}\sum_{\alpha_3\alpha_4}^{n}\sum_{\mu_5\neq\mu_6}^{m}e^{-2x\,(\mu_5+\mu_6)} 
\Dqq{1,1}{2,3}{1,2}{2,4}{3,5}{4,5}{3,6}{4,6}\Big)\nonumber\\ 
&\equiv&\frac{1}{Z^4}(A+2B)\,.\label{I2AB}
\end{eqnarray}
The terms $A$ and $B$ depend on 19 different diagrams which we
calculate again by invariant integration. For $A$ we have
\begin{eqnarray}
A&=&\sum_{\{\al_i,\mu_j\}}^{(n,m)}\sum_{\mu_5,\mu_7}^{m}e^{-2x\,(\mu_5+\mu_7)}
\Dqq{1,1}{1,2}{2,3}{2,4}{3,5}{4,5}{5,7}{6,7}\label{Adb1}\\ 
&=&\sum_{\{\al_i,\mu_j\}}^{(n,m)}\Big(\sum_{\mu_5=\mu_7}^{m}e^{-4x\,\mu_5}\Dqq{1,1}{1,2}{2,3}{2,4}{3,5}{4,5}{5,5}{6,5}+\sum_{\mu_5\neq\mu_7}^{m}e^{-2x\,(\mu_5+\mu_7)}\Dqq{1,1}{1,2}{2,3}{2,4}{3,5}{4,5}{5,7}{6,7}\Big)\label{Adb2}\\ 
&=&\sum_{\{\al_i,\mu_j\}}^{(n,m)}\Big(\Dqq{1,1}{1,2}{2,3}{2,4}{3}{4}{5}{6}(\sum_{\mu_5=\mu_7}^{m}e^{-4x\,\mu_5})+\Dqq{1,1}{1,2}{2,3}{2,4}{3}{4}{$\bar{5}$}{$\bar{6}$}(\sum_{\mu_5\neq\mu_7}^{m}e^{-2x\,(\mu_5+\mu_7)})\Big)\label{Adb3}\\ 
&=&\sum_{\{\al_i,\mu_j\}}^{(n,m)}\Big(f(x)\Dqq{1,1}{1,2}{2,3}{2,4}{3}{4}{5}{6}+g(x)\Dqq{1,1}{1,2}{2,3}{2,4}{3}{4}{$\bar{5}$}{$\bar{6}$}\Big)\,. \label{Adb4}
\end{eqnarray}
The single indices on the right hand sides of the diagrams in (\ref{Adb3})
now decode only $\alpha$'s; $\mu_1\ldots \mu_4$ still appear
explicitly on the left column of vertices, $\mu_6$ and $\mu_8$ were
chosen identical to $\mu_5$ and $\mu_7$, respectively, and the latter
two indices are still summed over.  A "bar" vertex is a vertex
which cannot collapse with a 
"normal" vertex even if both values of the corresponding $\alpha$'s
(or $\mu$'s) 
are the same.  
The vertices labeled $\bar{5}$ and $\bar{6}$ (which stand here for
$\alpha_5$ and 
$\alpha_6$) inherit this property from the $\mu_7$ part 
still present in (\ref{Adb2}): The restriction $\mu_5\ne\mu_7$ implies
indeed that none of the 
two top vertices can collapse with either of the two bottom vertices
in the second diagram.   Thus
"normal" and  "bar" vertices can only collapse on vertices of the
same kind. The constraint $\alpha_5\ne \alpha_6$ is still
implicit.
The functions $f(x)$ et $g(x)$ are defined as 
\begin{eqnarray}
f(x)&=&\left(\frac{1-e^{-4x\, d}}{e^{4x\,}-1}\right)=Z(4x)\label{f}\\
g(x)&=&e^{-6x\,}\left(\frac{1-e^{-2x\, d}}{1-e^{-2x\,}}\right)\left(\frac{1-e^{-2x\,(d-1)}}{1-e^{-4x\,}}\right)=Z^2(2x)-Z(4x)\,.\label{g}
\end{eqnarray}
With this, we have introduced all notational innovations which allow
the analytical calculation of $\bra \I^2\ket$. The rest of the calculations
amounts to identifying all possible non-zero 
configurations of collapsing vertices allowed by the remaining
summation variables. The explicit expansion of the terms $A$ and $B$ finally
leads to
\be \label{moy.I2}
\bra \I^2\ket=\frac{n}{Z^4}\Big[f(x)\Big(A_1+(n-1)A_3\Big) +
g(x)\Big(A_2+(n-1)A_4+n(n-1)B_1+n(n-1)^2B_2\Big) \Big]. 
\ee
Here, the parameter $d$ in the functions $f$ and $g$ is $d=m$. The terms
$A_i$ and $B_i$ are defined and calculated explicitly in the Appendix.
They only depend on $n$ and $m$. 
\begin{figure}[h]
\epsfig{file=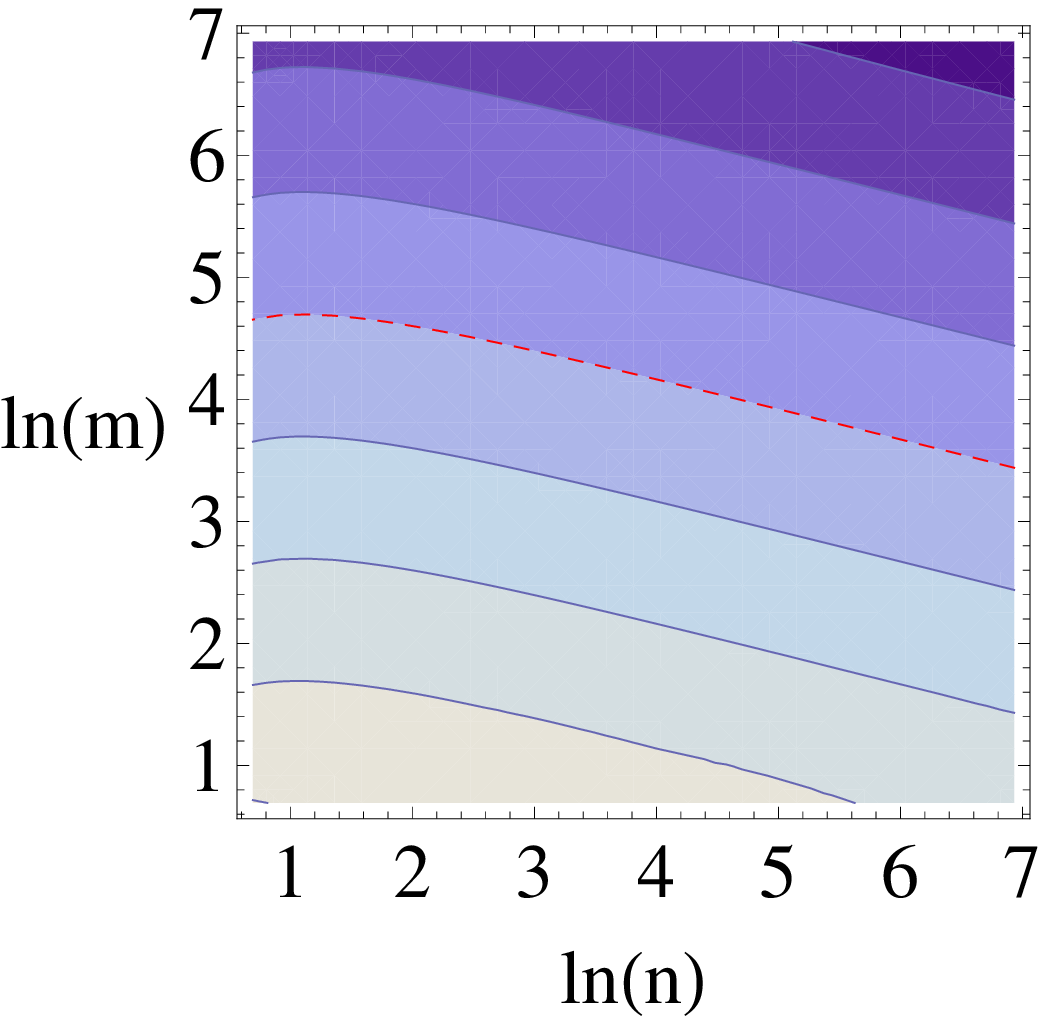,width=8cm,angle=0}
\epsfig{file=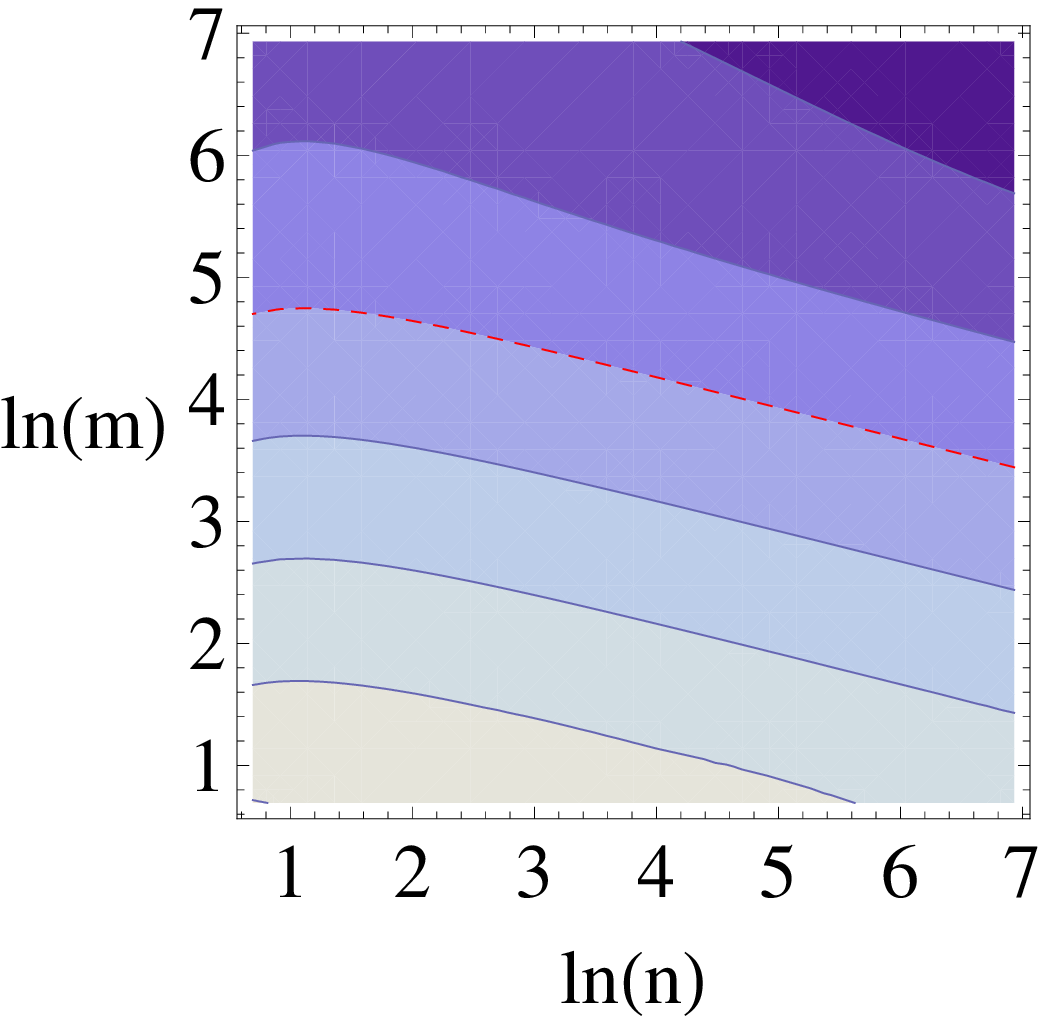,width=8cm,angle=0}\\
\epsfig{file=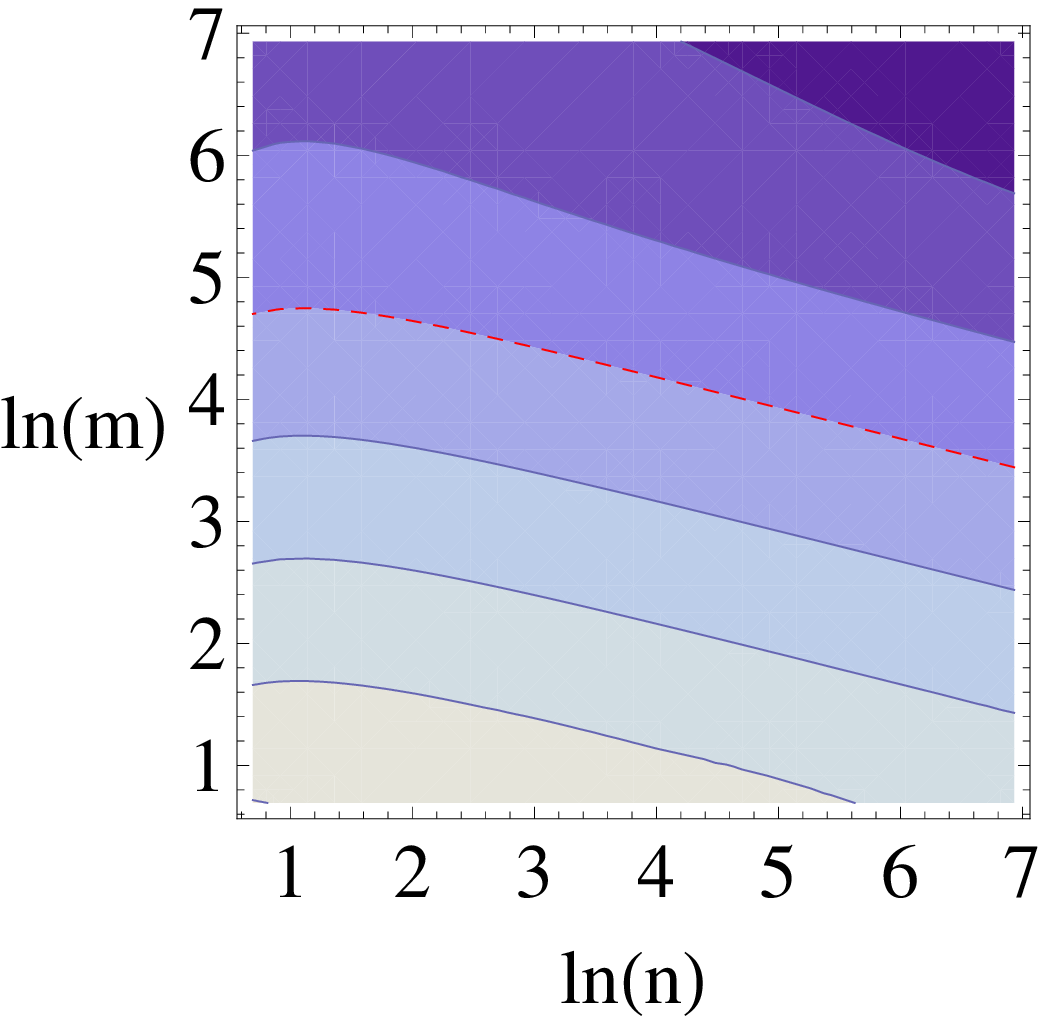,width=8cm,angle=0}
\epsfig{file=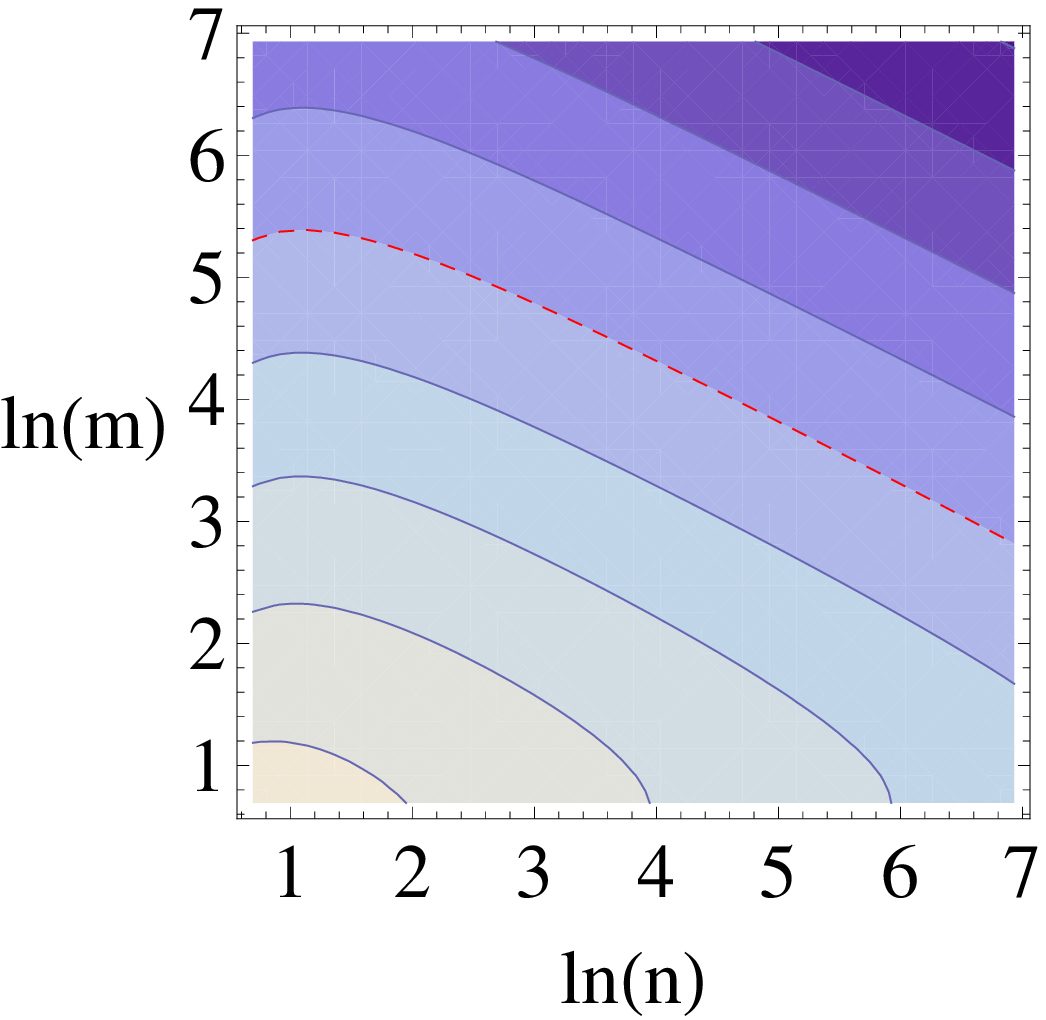,width=8cm,angle=0}
\caption{(Color online) Contour plot of $\ln(\sigma_{\I}(n,m))$ for x=0.001, 0.01, 0.1
  and 10 (upper left to lower right), for $n$ and $m$ between 2 and
  1024. The distance between the contours is 2 and the dashed line is for
  $\ln(\bra \I(n,m) \ket)=-10$, except for the last plot where the distance
  is 1 and the dashed line is for $\ln(\sigma_{\I}(n,m))=-6$. Values
  increase from  dark to bright colors. } 
\label{fig.sig} 
\end{figure} 

In Fig.\ref{fig.sig} we plot the standard deviation of the distribution of
$\I$, $\sigma_{\I}(n,m,x)=(\bra \I^2\ket - \bra \I \ket^2)^{1/2}$  for four
different temperatures as function of $n$ and $m$. For given temperature,
$\sigma_{\I}(n,m,x)$  decreases with $n$ and $m$. The log-log-log plot
shows that the decay behaves as a power law in $n$ and in $m$. The
corresponding powers can be found from an asymptotic expansion of the
variance $var(n,m,x)=\sigma_{\I}^{2}(n,m,x)$ for $n \gg 1$  or for $m \gg 1$
in the limits of zero or infinite 
temperature. For fixed $m$, we find for $n \gg 1$
\begin{eqnarray}
var(n,m,x\to\infty) &=& \frac{2(m-1)^2}{n m^4} - \frac{4(m^4-3m^3+3m^2-5m+3)}{m^6n^2} + O(\frac{1}{n^3})\label{varIxi} \\
var(n,m,x=0) &=& \frac{2(m^2-1)}{n m^6} + \frac{8-4m^4}{m^8n^2} + O(\frac{1}{n^3})\label{varIx0}\,.
\end{eqnarray}
This should be compared to the unitary case, where the asymptotic expansion
reads $var_{U}(n) = \frac{4}{n^2} + O(\frac{1}{n^3})$, as is still evident from eqs.(\ref{varIxi}) and (\ref{varIx0}) by choosing
$m=1$. We see
that the variance decays more slowly as function of $n$ in the presence of
decoherence, i.e.~as $1/n$ instead of as $1/n^2$ in the unitary case. In
other words,  decoherence tends to slow down convergence of the interference distribution to a narrow peak.
Nevertheless, the power law decay of the variance as function of $n$
implies that, also in the non-unitary case, the interference distribution
becomes for $n \gg 1$ a very narrow peak centered about the average value
(which itself 
increases with $n$, see eqs.(\ref{avIxInfty}) and (\ref{avIx0})). 

Asymptotic expansion of  $var(n,m,x)$ as function of $m\gg 1$ with fixed $n$ 
gives 
\begin{eqnarray}
var(n,m,x\to\infty) &=& \frac{2(n-1)^2}{n^3m^2} + O(\frac{1}{m^3})\\
var(n,m,x=0) &=& \frac{(n-1)^2}{n^3 m^4} + O(\frac{1}{m^6})\,.
\end{eqnarray}
Thus, also  an increase of the dimension of the environmental Hilbert
space narrows the interference distribution.  However, since according to (\ref{moyIxtoinf},\ref{moyIxto0}),  the average
interference decays as $1/m$ ($1/m^2$) for $x\to\infty$ ($x\to 0$), the
relative width, i.e.~standard deviation divided by the average value, is
asymptotically independent of the dimension of the environment.

In the case $m=d=1$ ($n=N$) all the prefactors $m[i]$ (see Appendix) are zero if $i\ge1$. With the same parameters we have furthermore from eqs.(\ref{f},\ref{g}), $f(x)=Z=1$, and $g(x)=0$. Thus the expression (\ref{moy.I2}) of $\bra \I^2 \ket$ simplifies
considerably, 
\begin{eqnarray*}
\bra \I^2 \ket&=&\Big(N A_{1}+N(N-1)A_{3}\Big)\\
&=&\Big(N\big(N[3]\DDuq+4N[2]\DDut+2N[1]\DDud\big)\\
& &+N(N-1)\big(N[3]\DDadq+4N[2]\DDadt+2N[1]\DDadd\big)\Big)\\
&=&\frac{N \left(N^3-5 N+8\right)-4}{(N+1) (N+3)}\,.
\end{eqnarray*}
As expected this leads to the standard deviation $\sigma_{\I}=\frac{2}{N+1}\sqrt{\frac{N-1}{N+3}}$, identical to the
expression for purely unitary propagation \cite{Arnaud07}.\\

The numerical results in section \ref{sec.numerical} are in very good
agreement with our analytical results, as can be seen in table \ref{tab.1}
where we compare the 
numerically obtained average values and standard deviations for the examples
shown in fig.\ref{fig.num2} and for $(n,m)=(4,8)$ and (8,4) to the
corresponding analytical results. 
\begin{table}[h]
\begin{center}
\begin{tabular}{|c|c|c|c|c|c|}\hline
n&m&$\bra \cI\ket$ (num.)&$\bra
\cI\ket$ (ana.)&$\sigma_\cI$ (num.) &$\sigma_\cI$ (ana.) \\\hline
4 & 2 & 0.57279 & 0.57286 & 0.11728 & 0.11719 \\
4 & 4 & 0.14296 & 0.14293 & 0.03260 & 0.03255 \\
4 & 8 & 0.03702 & 0.03702 & 0.00864 & 0.00864 \\
8 & 2 & 1.54120 & 1.54109 & 0.09022 & 0.09409 \\  
8 & 4 & 0.38796 & 0.38796 & 0.02670 & 0.02666 \\  
\hline\end{tabular}
\caption{Comparison of numerical and analytical values of $\bra \cI\ket$ and
  $\sigma_\cI$. All results are rounded to five digits after the decimal
  point.   \label{tab.1}}  
\end{center}
\end{table}

\section{Interference for a spin coupled to  several spins} 

In this part, we generalize the previous calculations to a situation where
the environment consists of $s$ independent spins with $d$ energy levels
with energy 
spacing $\hbar\Omega$. Thus, the dimension of the environment is $m=d^s$. The 
hamiltonian of this system reads
\be
H^{(s)}=\sum_{k=1}^{s}H_{k}^{(1)}\,,
\ee
where $H_k^{(1)}$ is the hamiltonian of spin number $k$ (eq.(\ref{h1})). The
components of $H^{(s)}$ in its eigenbasis are
\begin{eqnarray}
H^{(s)}_{\nu \rho}&=&\hbar\Omega\left(\sum_{k=1}^{s}\nu_k\right)\delta_{\nu \rho}
\end{eqnarray}
with the notation for the indices $\nu=(\nu_1,\nu_2,...,\nu_s)$ and $\rho=(\rho_1,\rho_2,...,\rho_s)$\\

The density matrix corresponding to the thermal state of such a system
factorizes, 
$
\epsilon^{(s)}=\epsilon^{(s)}=\epsilon^{(1)\otimes s},
$
which leads to the components
\begin{eqnarray}
\epsilon^{(s)}_{\nu \rho}&=&\frac{e^{-x\,S(\nu)}\delta_{\nu \rho}}{Z^s}
\end{eqnarray}
with $x=\beta\hbar\Omega$, $S(\nu)=\sum_{k=1}^{s}\nu_k$, and where $Z$ is the
partition function of the thermal state of a single spin introduced
in the previous section.  It turns out that in order to generalize the
previous 
calculation of $<\I>$ and $<\I^2>$ to this kind of environment, we just have to replace $Z$ by $Z^s$ in eqs.(\ref{Iavfin}), (\ref{f}), and (\ref{g}), and
keep $d=m^s$ 
instead of $d=m$. This is again a consequence of the fact that the values of
the diagrams do not depend on the indices of the vertices.  Thus, the same
values are obtained even for composite indices reflecting
several subsystems, and only the
multiplicities and temperature dependent factors are modified. 
Since the spins of the heat bath are taken as non-interacting, the
sums over the thermal factors just gives rise to powers of the single spin
thermal factors, as is the case also for the calculation of the partition
function for $s$ spins. This means that we have to replace
\begin{eqnarray*}
f(x)&\rightarrow&\sum_{\mu}e^{-4xS(\mu)}=\sum_{\mu_{1}}^{d}...\sum_{\mu_{s}}^{d}e^{-4x\mu_{1}}...e^{-4x\mu_{s}}=Z^s(4x)=f^{s}(x)\\ 
g(x)&\rightarrow&\sum_{\mu\neq\nu}e^{-2x(S(\mu)+S(\nu))}=\sum_{\mu,\nu}e^{-2x(S(\mu)+S(\nu))}-\sum_{\mu}e^{-4xS(\mu)}=Z^s(2x)-Z^s(4x)\,.
\end{eqnarray*}
The expressions for $\bra \I \ket$ and $\bra \I^2 \ket$ become
\begin{eqnarray}
\bra \I \ket&=& n^{2}(n-1)h^s(x)\Big(d\FFuu+d(d-1)\EEd\Big)\label{1}\\
&=&\Big(
\coth(\frac{dx}{2})\tanh(\frac{x}{2})\Big)^{s}\Big(\frac{nd^s(n-1)^2}{n^2d^{2s}-1}\Big)\,,\\ 
\bra \I^{2}
\ket&=&\frac{n}{Z^{4s}}\Big[f^{s}(x)\Big(A_1+(n-1)A_3\Big)\nonumber\\ 
&& + g^{s}(x)\Big(A_2+(n-1)A_4+n(n-1)B_1+n(n-1)^2B_2\Big) \Big]\,.\label{2}
\end{eqnarray}
The argument $m$ in the terms $A_i$ and $B_i$ in eqs.(\ref{1},\ref{2}) is
now $m = d^s$.   It means that $s$ spins of size $(d-1)/2$ act very similarly
as a single spin of size $(d^s-1)/2$, when it comes to
their influence on the first and 
second moments of $P(\cI)$.  The only difference lies in the temperature
dependent prefactors $f(x),g(x)$ and $h(x)$. For a single spin  of size
$(d^s-1)/2$, $d$ in eqs.(\ref{h},\ref{f},\ref{g}) is given by the dimension
of the environment $m=d^s$, but 
in eqs.(\ref{1},\ref{2}) we have $s=1$ for a single spin. For $s$ spins of
size $(d-1)/2$  the 
dimension $d$ in eqs.(\ref{h},\ref{f},\ref{g}) remains, and $s$ is the
number of spins in eqs. (\ref{1},\ref{2}). In the limits $x\to 0$ or
$x\to\infty$ the expressions coincide for the two situations. 

\section{Summary}
We have investigated quantitatively how quantum interference is
affected by decoherence.  Based on a distribution of unitary matrices drawn
from 
CUE which describe the joint propagation of system and heat bath, we have
shown that the average interference increases roughly linearly with the
Hilbert space dimension $n$ of the system, but decays as a power of
the dimension $m$ of the environment. That power depends on the temperature
of the environment (chosen here as one or several non-interacting spins),
with a decay that 
essentially scales like $1/m^2$ for $T=0$, and as $1/m^3$ for
$T\to\infty$.  The width of the distribution decreases more slowly when
decoherence 
becomes important, but for fixed $m$, the width of the distribution still
decays as $1/\sqrt{n}$ (instead of as $1/n$ in the unitary case).  Thus,
for $n\gg 1$ and $m$ fixed, the distribution of quantum interference is
still a sharp 
peak concentrated on the average value.  Numerically we have shown that the
interference distribution in the non-unitary case can be well fitted to a
log-normal distribution for sufficiently large $n$, which implies that the
number of i-bits \cite{Braun06} is to good approximation Gaussian distributed. 

{\em Acknowledgments:}
We would like to thank 
CALMIP (Toulouse) for the use of their computers. This
work was supported by the Agence  
National de la Recherche (ANR), project INFOSYSQQ. 

\section{Appendix}
We provide here the remaining details of the calculation of the terms
$A$ and $B$ in the expression for $\bra \I^2\ket$, eq.(\ref{I2AB}), as
well as the values of the resulting diagrams.

\subsection{The $A$ term}
>From eq.(\ref{Adb4}) we have 
\begin{eqnarray}\label{A}
A&=&\sum_{\{\al_i,\mu_j\}}^{(n,m)}\Big(\sum_{\{\al_1=\al_2\}}^{n}f(x)\Dqq{1,1}{1,2}{1,3}{1,4}{3}{4}{5}{6}+g(x)\Dqq{1,1}{1,2}{1,3}{1,4}{3}{4}{$\bar{5}$}{$\bar{6}$}\nonumber\\
& &+\sum_{\{\al_1\neq\al_2\}}^{n}f(x)\Dqq{1,1}{1,2}{2,3}{2,4}{3}{4}{5}{6}+g(x)\Dqq{1,1}{1,2}{2,3}{2,4}{3}{4}{$\bar{5}$}{$\bar{6}$}\Big)\nonumber\\
&=&\sum_{\{\al_i,\mu_j\}}^{(n,m)}\Big(n\,f(x)\Dqq{1}{2}{3}{4}{3}{4}{5}{6}+n\,g(x)\Dqq{1}{2}{3}{4}{3}{4}{$\bar{5}$}{$\bar{6}$}\nonumber\\
& &+n(n-1)\,f(x)\Dqq{1}{2}{$\bar{3}$}{$\bar{4}$}{3}{4}{5}{6}+n(n-1)\,g(x)\Dqq{1}{2}{$\bar{3}$}{$\bar{4}$}{3}{4}{$\bar{5}$}{$\bar{6}$}\Big)\nonumber\\
&=&n\,f(x)A_1+n\,g(x)A_2+n(n-1)\,f(x)A_3+n(n-1)\,g(x)A_4\,.
\end{eqnarray}
By taking into account the constraints on the $\alpha_i$ we get
\begin{eqnarray*}
A_1&=&\sum_{\{\al_i,\mu_j\}}^{(n,m)}\Dqq{1}{2}{3}{4}{3}{4}{5}{6}\\
&=&\sum_{\{\mu_j\}}^{m}\left(n[3]\Dqq{1}{2}{3}{4}{}{}{}{} + 4n[2]\Dqt{1}{2}{3}{4}{}{}{} + 2n[1]\Dqd{1}{2}{3}{4}{}{}\right)\\
&=&n[3]A_{11}+4n[2]A_{12}+2n[1]A_{13}\,,
\end{eqnarray*}
with $n[i]=n(n-1)(n-2)...(n-i)$. We check that we have the
$n[3]+4n[2]+2n[1]=n^{2}(n-1)^{2}$ configurations corresponding to the
sum over the four indices $\alpha_j$ with the two constrains
$\alpha_3\neq\alpha_4$ and $\alpha_5\neq\alpha_6$.
The $A_{1k}$ read
\begin{eqnarray*}
A_{11}&=&\sum_{\{\mu_j\}}^{m}\Dqq{1}{2}{3}{4}{}{}{}{}\\
&=&m[3]\DDqq+m[2]\Big(4\DDatq+2\DDbtq\Big)\\
& &+m[1]\Big(2\DDcdq+\DDadq+4\DDbdq\big)+m\DDuq,\\
A_{12}&=&\sum_{\{\mu_j\}}^{m}\Dqt{1}{2}{3}{4}{}{}{}\\
&=&m[3]\DDqt+2m[2]\Big(\DDbtt+\DDctt+\DDatt\Big)\\
& &+m[1]\Big(\DDadt+\DDddt+\DDcdt \Big)+4m[1]\DDbdt+m\DDut,\\
A_{13}&=&\sum_{\{\mu_j\}}^{m}\Dqd{1}{2}{3}{4}{}{}\\
&=&m[3]\DDqd+m[2]\Big(4\DDatd+2\DDbtd\Big)\\
& &+m[1]\Big(2\DDadd+\DDbdd+4\DDcdd\Big)+m\DDud\,.
\end{eqnarray*}
For $A_2$ we obtain directly
\begin{eqnarray*}
A_2&=&\sum_{\{\al_i,\mu_j\}}^{(n,m)}\Dqq{1}{2}{3}{4}{3}{4}{$\bar{5}$}{$\bar{6}$}=n^2(n-1)^2A_{11}\,,
\end{eqnarray*}
whereas $A_3$ is given by 
\begin{eqnarray*}
A_3&=&\sum_{\{\al_i,\mu_j\}}^{(n,m)}\Dqq{1}{2}{$\bar{3}$}{$\bar{4}$}{3}{4}{5}{6}\\
&=&\sum_{\{\mu_j\}}^{m}\left(n[3]\Dqq{1}{2}{$\bar{3}$}{$\bar{4}$}{}{}{}{} + 4n[2]\Dqt{1}{2}{$\bar{3}$}{$\bar{4}$}{}{}{} + 2n[1]\Dqd{1}{2}{$\bar{3}$}{$\bar{4}$}{}{}\right)\\
&=&n[3]A_{31}+4n[2]A_{32}+2n[1]A_{33}\,.
\end{eqnarray*}
The $A_{3k}$ are
\begin{eqnarray*}
A_{31}&=&\sum_{\{\mu_j\}}^{m}\Dqq{1}{2}{$\bar{3}$}{$\bar{4}$}{}{}{}{}\\
&=&\Big(m[3]+4m[2]+2m[1]\Big)\DDqq+\Big(2m[2]+4m[1]\Big)\DDbtq+\Big(m[1]+m\Big)\DDadq\,\\
A_{32}&=&\sum_{\{\mu_j\}}^{m}\Dqt{1}{2}{$\bar{3}$}{$\bar{4}$}{}{}{}\\
&=&\Big(m[3]+4m[2]+2m[1]\Big)\DDqt+(2m[2]+4m[1])\DDbtt+(m[1]+m)\DDadt\,\\
A_{33}&=&\sum_{\{\mu_j\}}^{m}\Dqd{1}{2}{$\bar{3}$}{$\bar{4}$}{}{}\\
&=&\Big(m[3]+4m[2]+2m[1]\Big)\DDqd+\Big(2m[2]+4m[1]\Big)\DDatd+\Big(m[1]+m\Big)\DDadd\,.\\
\end{eqnarray*}
The term $A_4$ can be expressed in terms of $A_{31}$,
\begin{eqnarray*}
A_4&=&\sum_{\{\al_i,\mu_j\}}^{(n,m)}\Dqq{1}{2}{$\bar{3}$}{$\bar{4}$}{3}{4}{$\bar{5}$}{$\bar{6}$}=n^2(n-1)^2A_{31}\,.
\end{eqnarray*}

As a consistency check, we verify in the calculation of the terms
$A_{3i}$ that we have the $m[3]+6m[2]+7m[1]+m=m^4$ configurations 
corresponding to the sum over the four indices $\mu_j$.  

\subsection{The $B$ term}
In the same way as for $A$, we find for the $B$ term
\begin{eqnarray}\label{B}
B&=&\sum_{\{\al_i,\mu_j\}}^{(n,m)}\sum_{\alpha_3,\alpha_4}^{n}\sum_{\mu_5\neq\mu_6}^{m}e^{-2x\,(\mu_5+\mu_6)} \Dqq{1,1}{2,3}{1,2}{2,4}{3,5}{4,5}{3,6}{4,6}\nonumber \\
&=&\sum_{\{\al_i,\mu_j\}}^{(n,m)}\sum_{\alpha_3,\alpha_4}^{n}\Big(g(x)\Dqq{1,1}{2,3}{1,2}{2,4}{3}{4}{$\bar{3}$}{$\bar{4}$}\Big) \nonumber\\
&=&n(n-1)g(x)\sum_{\{\al_i,\mu_j\}}^{(n,m)}\Dqq{1,1}{2,3}{1,2}{2,4}{}{}{}{}\nonumber \\
&=&n(n-1)g(x)\sum_{\{\mu_j\}}^{m}\sum_{\al_1\al_2}^{n}\Big(g(x)\Dqq{1,1}{2,3}{1,2}{2,4}{}{}{}{}\Big)\nonumber \\
&=&n(n-1)g(x)\sum_{\{\mu_j\}}^{m}\Big(n\,\Dqq{1}{3}{2}{4}{}{}{}{}+n(n-1)\,\Dqq{1}{$\bar{3}$}{2}{$\bar{4}$}{}{}{}{}\Big)\nonumber \\
&=&n^{2}(n-1)g(x)\Big(B_1+(n-1)B_2\Big)\,,
\end{eqnarray}
where the terms $B_i$ are given by 
\begin{eqnarray*}
B_1&=&\sum_{\{\mu_j\}}^{m}\Dqq{1}{2}{3}{4}{}{}{}{}=A_{11}\\
&=&m[3]\DDqq+4m[2]\DDatq+2m[2]\DDbtq\\
& &+m[1]\left(\DDadq+2\DDcdq+4\DDbdq\right)+m\DDuq,\\
B_2&=&\sum_{\{\mu_j\}}^{m}\Dqq{1}{$\bar{2}$}{3}{$\bar{4}$}{}{}{}{}\\
&=&\Big(m[3]+4m[2]+2m[1]\Big)\DDqq+\Big(2m[2]+4m[1]\Big)\DDatq+\Big(m[1]+m\Big)\DDcdq\,.
\end{eqnarray*}

\subsection{Analytical expressions for all diagrams}

All diagrams can be calculated by invariant integration. We find
\begin{tabular}{lr}
$\FFuu=\frac{1}{N(N + 1)}$ &
$\EEd=\frac{-1}{N(N^2-1)}$\\
$\DDut=\frac{2}{(N + 3) (N + 2) (N + 1) N}$ &
$\DDuq=\frac{1}{(N + 3) (N + 2) (N + 1) N}$\\
$\DDadd=\frac{N^2 + N + 2}{(N + 3) (N + 2) (N^2 - 1) N^2}$ &
$\DDbdd=\frac{8}{(N + 3) (N + 2) (N^2 - 1) N^2}$\\
$\DDcdd=\frac{-4}{(N + 3)(N + 2) (N^2 - 1) N}$ &
$\DDadt=\frac{N + 1}{(N + 3) (N + 2) N^2 (N - 1)}$\\
$\DDbdt=\frac{-2}{(N + 3) (N + 2) (N^2 - 1) N}$ &
$\DDcdt=\frac{-1}{(N + 3) (N + 2) (N + 1) N^2}$\\
$\DDadq=\frac{1}{(N + 3) (N - 1) N^2}$ &
$\DDbdq=\frac{-1}{(N + 3) (N + 2) (N^2 - 1) N}$\\
$\DDcdq=\frac{2}{(N + 3) (N + 2) (N^2 - 1) N^2}$ &
$\DDatd=\frac{-1}{(N + 3) (N + 2) (N + 1) N^2}$\\
$\DDbtd=\frac{4}{(N + 3) (N + 2) (N^2 - 1) N^2}$&
$\DDatt=\frac{3 N - 1}{(N + 3) (N^2 - 4) (N^2 - 1) N^2}$\\
$\DDbtt=\frac{-(N^2 + 1)}{(N + 3) (N^2 - 4) (N^2 - 1) N^2}$ &
$\DDctt=\frac{2}{(N + 3) (N + 2) (N^2 - 1) N^2}$\\
$\DDatq=\frac{1}{(N + 3) (N + 2) (N^2 - 1) N^2}$ &
$\DDbtq=\frac{-(N^2 + 2 N + 2)}{(N + 3) (N^2 - 4) (N^2 - 1) N^2}$\\
$\DDqd=\frac{2}{(N + 3)(N + 2)(N^2 - 1) N^2}$ &
$\DDqt=\frac{1}{(N + 3)(N + 2)(N^2 - 1) N^2}$\\
$\DDqq=\frac{(N^2 + 6)}{(N^2 - 9) (N^2 - 4) (N^2 - 1) N^2}$\,. & 
\end{tabular}

\bibliography{../mybibs_bt}

\end{document}

%% file: diagramsLA.tex
\newcommand{\Fuu}[3]{{\scriptsize \begin{tabular}{c} #1 \end{tabular} \begin{tabular}{c} \raisebox{-0.4cm}{\includegraphics[width=1.2cm]{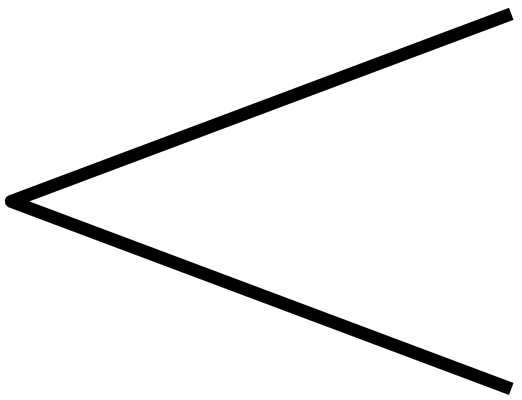}} \end{tabular} \begin{tabular}{c} #3\\ \\#2 \end{tabular}}}

\newcommand{\Ed}[4]{{\scriptsize \begin{tabular}{c} #2\\#1 \end{tabular} \begin{tabular}{c} \raisebox{-0.4cm}{\includegraphics[width=1.2cm]{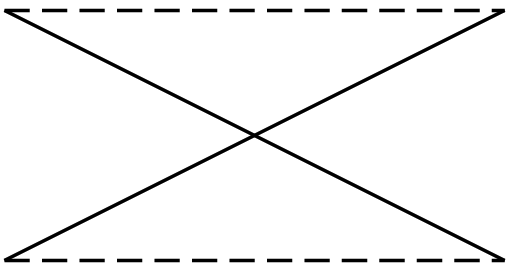}} \end{tabular} \begin{tabular}{c} #4\\#3 \end{tabular}}}

\newcommand{\Fmax}[6]{{\scriptsize \begin{tabular}{c} \\#2\\ \\#1 \end{tabular} \begin{tabular}{c} \raisebox{-1.3cm}{\includegraphics[width=2.1cm]{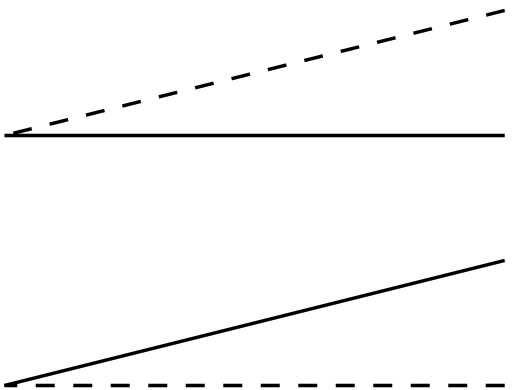}} \end{tabular} \begin{tabular}{c} #6\\#5\\#4\\#3 \end{tabular}}}

\newcommand{\Dqd}[6]{{\scriptsize \begin{tabular}{c} #4\\#3\\#2\\#1 \end{tabular} \begin{tabular}{c} \includegraphics[width=1cm]{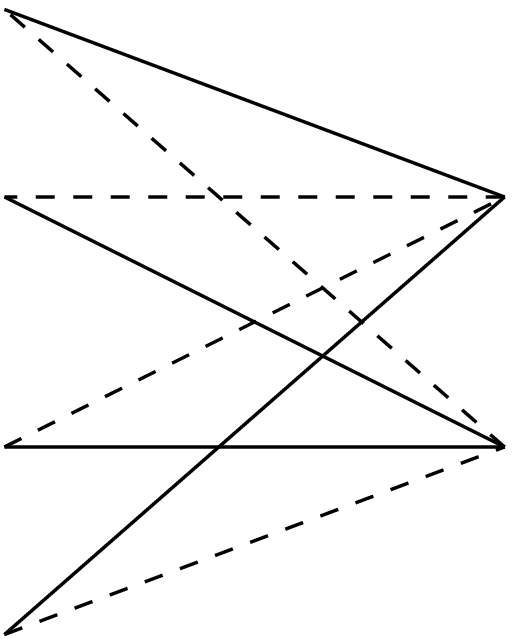} \end{tabular} \begin{tabular}{c} #6\\#5 \end{tabular}}}

\newcommand{\Dqt}[7]{{\scriptsize \begin{tabular}{c} #4\\#3\\#2\\#1 \end{tabular} \begin{tabular}{c} \includegraphics[width=1cm]{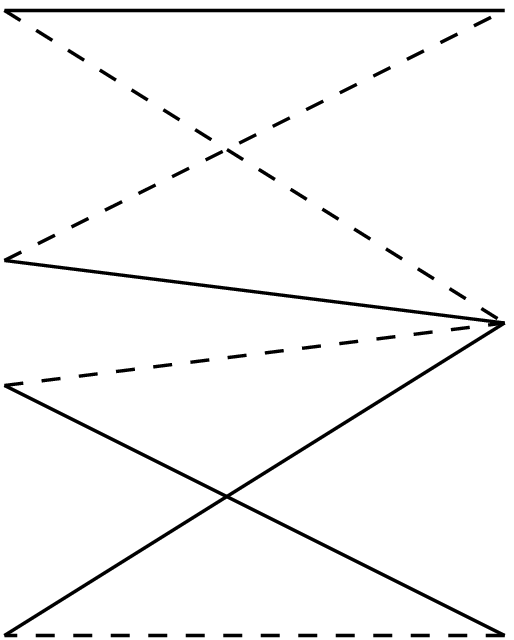} \end{tabular} \begin{tabular}{c} #7\\#6\\#5 \end{tabular}}}

\newcommand{\Dqq}[8]{{\scriptsize \begin{tabular}{c} #4\\#3\\#2\\#1 \end{tabular} \begin{tabular}{c} \raisebox{-1.3cm}{\includegraphics[width=1.3cm]{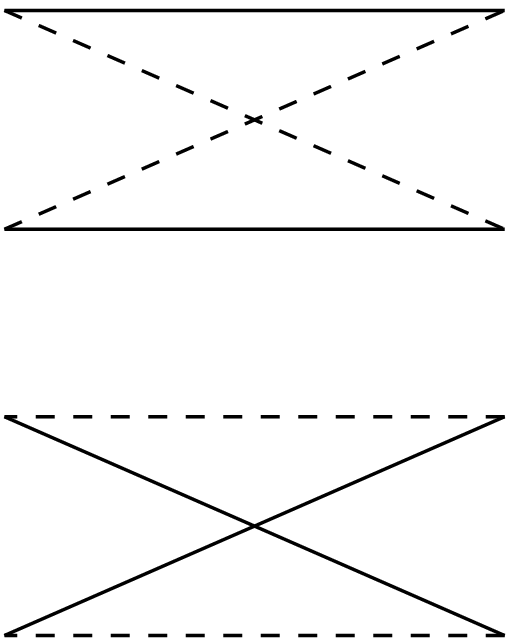}} \end{tabular} \begin{tabular}{c} #8\\#7\\#6\\#5 \end{tabular}}}

\newcommand{\Dmax}{{\scriptsize \begin{tabular}{c} \\2,4\\ \\2,3\\ \\1,2\\ \\1,1 \end{tabular} \begin{tabular}{c} \raisebox{-0.8cm}{\includegraphics[width=2.1cm]{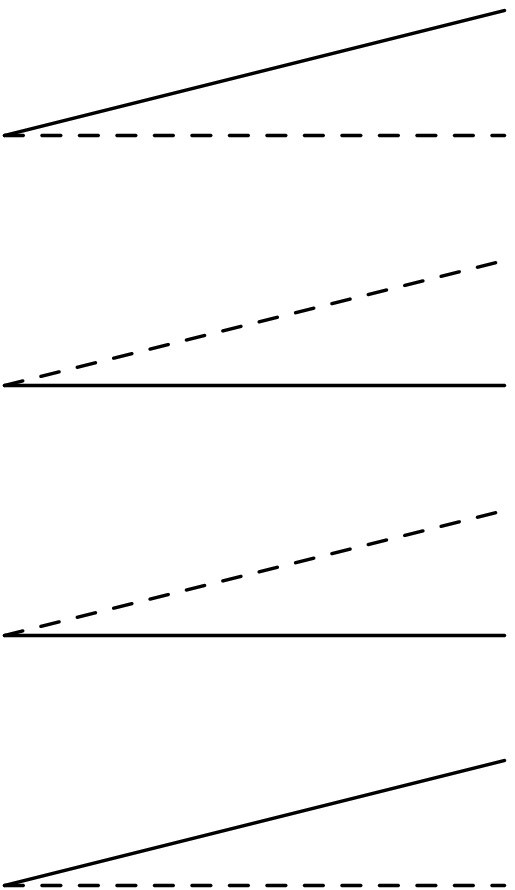}} \end{tabular} \begin{tabular}{c} 6,8\\5,8\\6,7\\5,7\\4,6\\3,6\\4,5\\3,5 \end{tabular}}}

\newcommand{\FFuu}{\begin{tabular}{c} \raisebox{-0.2cm}{\includegraphics[width=1.5cm]{diag/F11.eps}} \end{tabular}}

\newcommand{\EEd}{\begin{tabular}{c} \raisebox{-0.2cm}{\includegraphics[width=1.5cm]{diag/E2.eps}} \end{tabular}}

\newcommand{\FFmax}{\begin{tabular}{c} \raisebox{0cm}{\includegraphics[width=1.5cm]{diag/Fmax.eps}} \end{tabular}}

\newcommand{\DDud}{\begin{tabular}{c} \includegraphics[width=1cm]{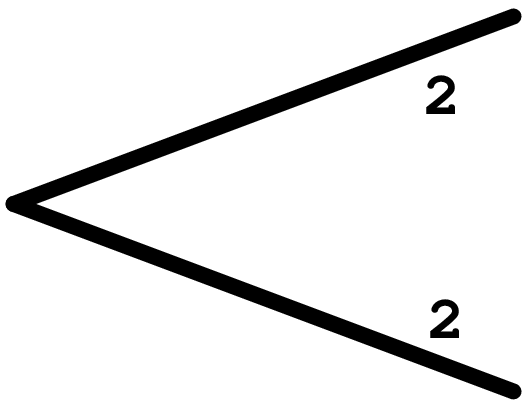} \end{tabular}}

\newcommand{\DDut}{\begin{tabular}{c} \includegraphics[width=1cm]{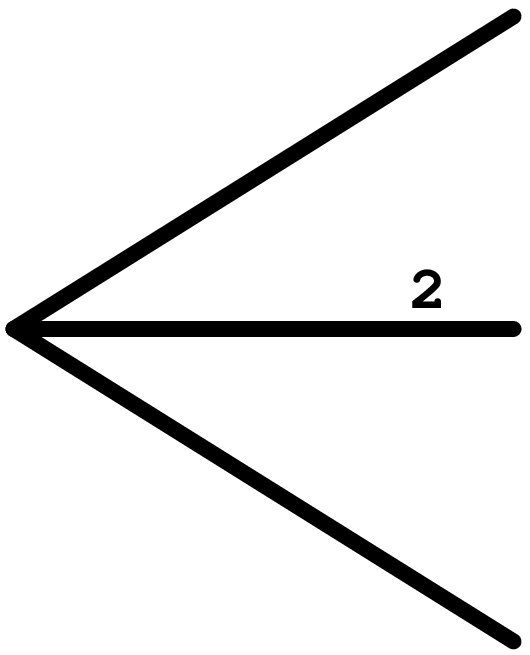} \end{tabular}}

\newcommand{\DDuq}{\begin{tabular}{c} \includegraphics[width=1cm]{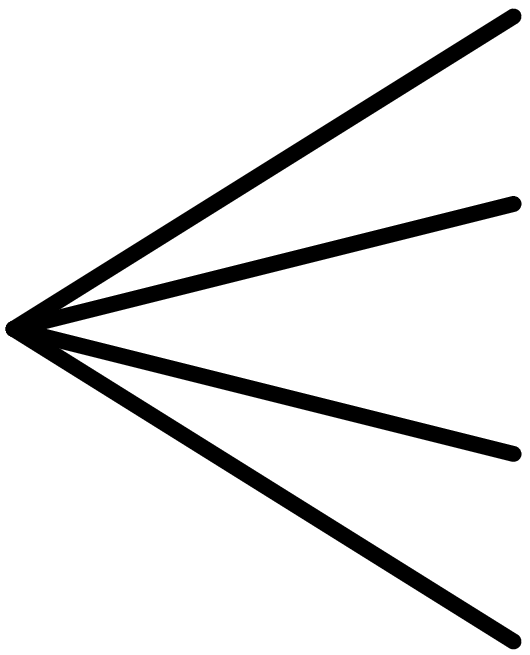} \end{tabular}}

\newcommand{\DDadd}{\begin{tabular}{c} \includegraphics[width=1cm]{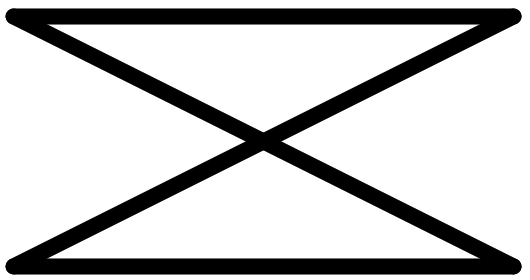} \end{tabular}}

\newcommand{\DDbdd}{\begin{tabular}{c} \includegraphics[width=1cm]{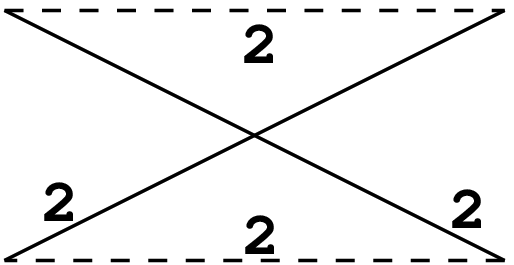} \end{tabular}}

\newcommand{\DDcdd}{\begin{tabular}{c} \includegraphics[width=1cm]{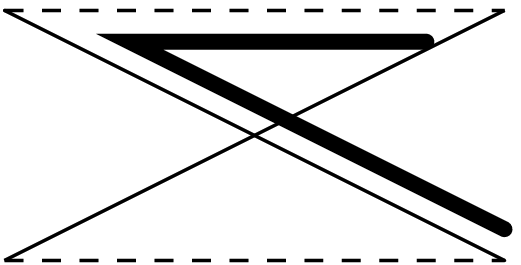} \end{tabular}}

\newcommand{\DDadt}{\begin{tabular}{c} \includegraphics[width=1cm]{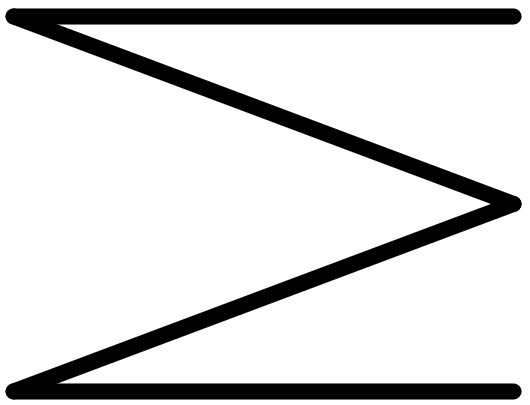} \end{tabular}}

\newcommand{\DDbdt}{\begin{tabular}{c} \includegraphics[width=1cm]{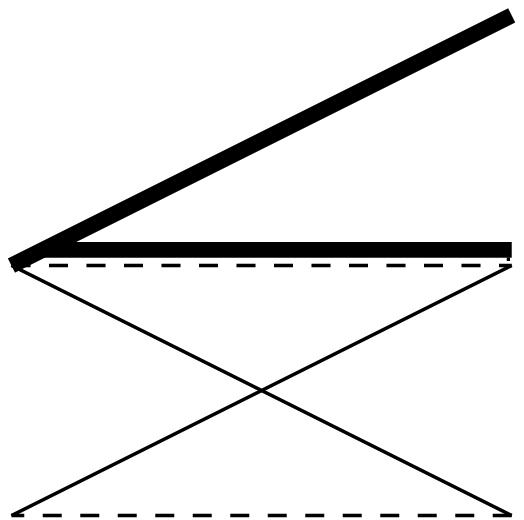} \end{tabular}}

\newcommand{\DDcdt}{\begin{tabular}{c} \includegraphics[width=1cm]{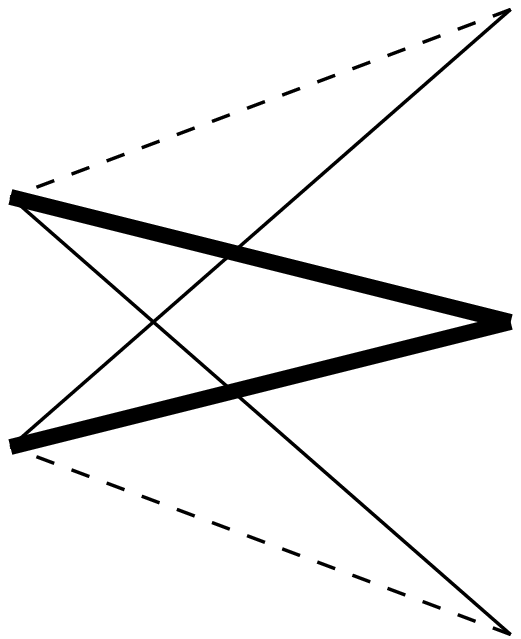} \end{tabular}}

\newcommand{\DDddt}{\begin{tabular}{c} \includegraphics[width=1cm]{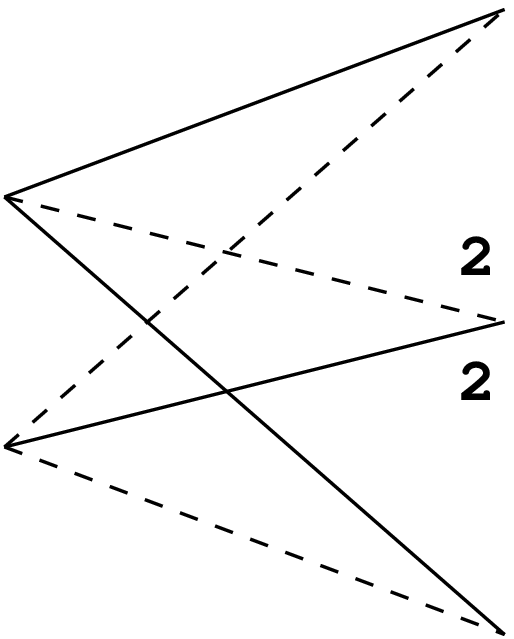} \end{tabular}}

\newcommand{\DDadq}{\begin{tabular}{c} \includegraphics[width=1cm]{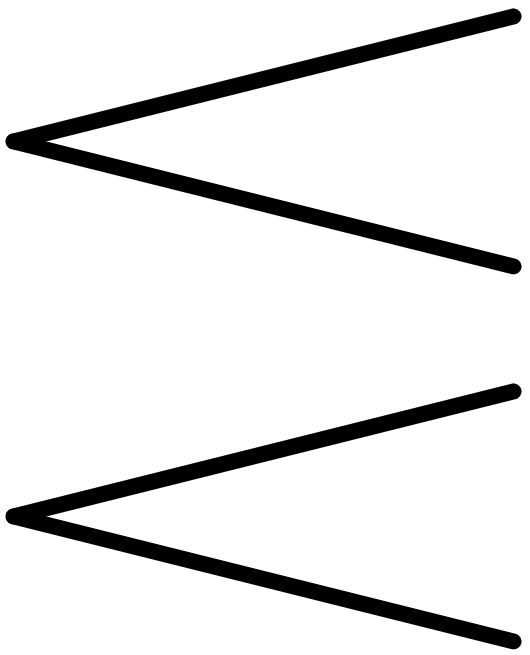} \end{tabular}}

\newcommand{\DDbdq}{\begin{tabular}{c} \includegraphics[width=1cm]{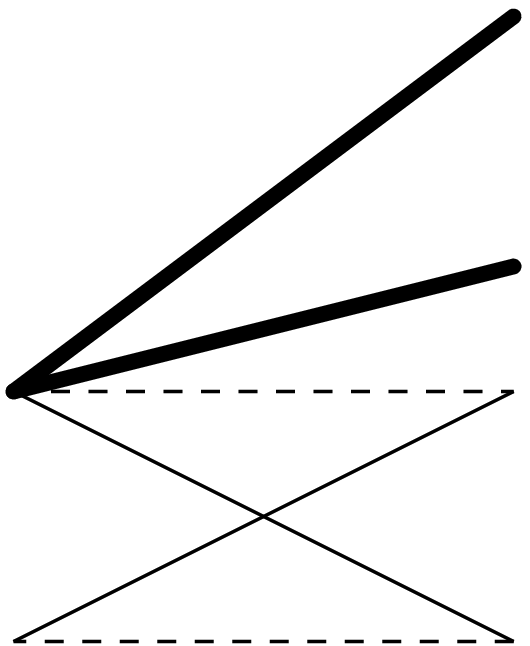} \end{tabular}}

\newcommand{\DDcdq}{\begin{tabular}{c} \includegraphics[width=1cm]{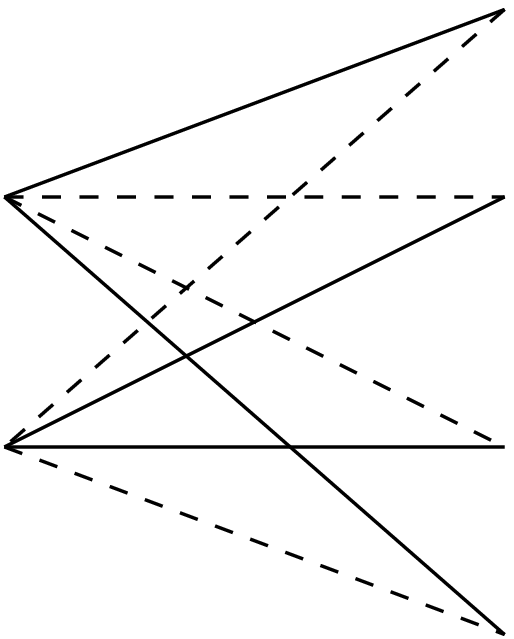} \end{tabular}}

\newcommand{\DDatd}{\begin{tabular}{c} \includegraphics[width=1cm]{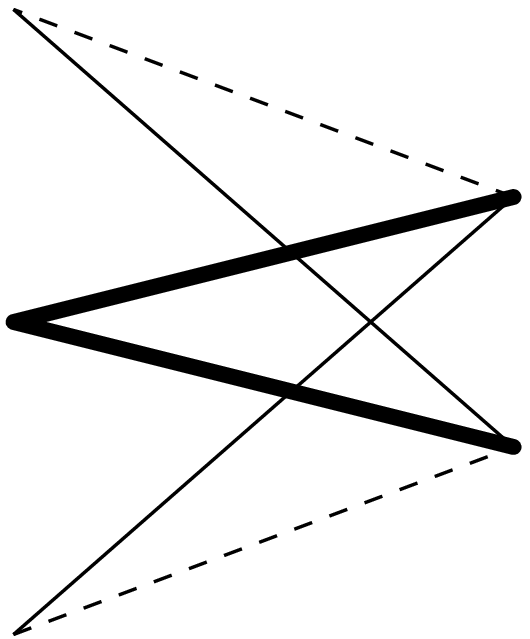} \end{tabular}}

\newcommand{\DDbtd}{\begin{tabular}{c} \includegraphics[width=1cm]{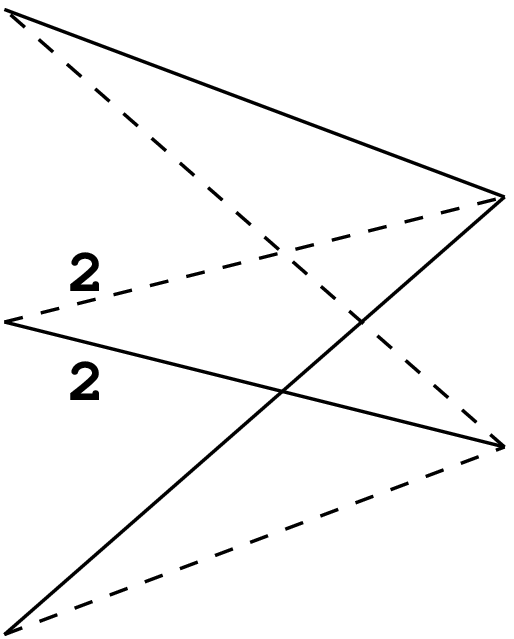} \end{tabular}}

\newcommand{\DDatt}{\begin{tabular}{c} \includegraphics[width=1cm]{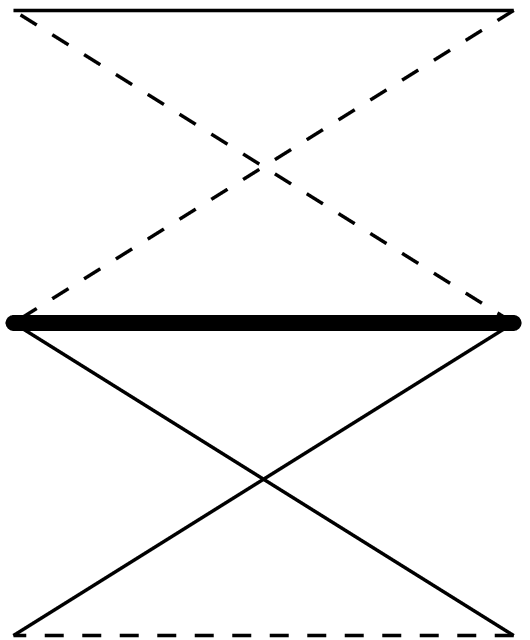} \end{tabular}}

\newcommand{\DDbtt}{\begin{tabular}{c} \includegraphics[width=1cm]{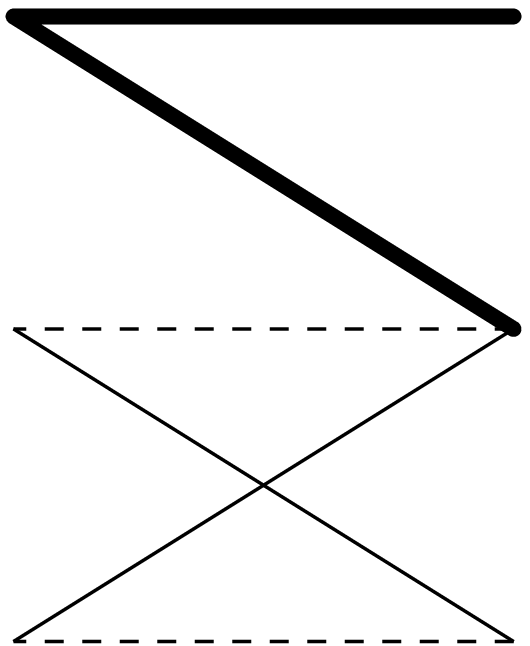} \end{tabular}}

\newcommand{\DDctt}{\begin{tabular}{c} \includegraphics[width=1cm]{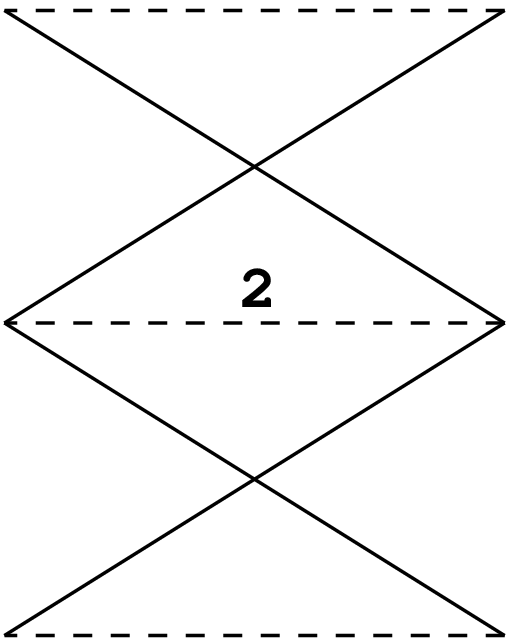} \end{tabular}}

\newcommand{\DDatq}{\begin{tabular}{c} \includegraphics[width=1cm]{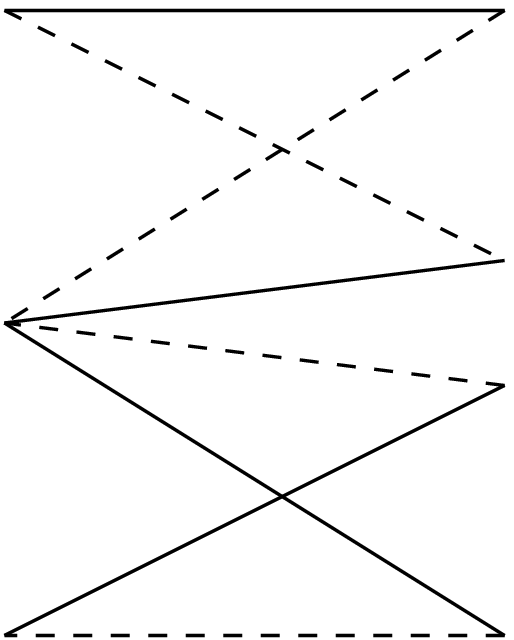} \end{tabular}}

\newcommand{\DDbtq}{\begin{tabular}{c} \includegraphics[width=1cm]{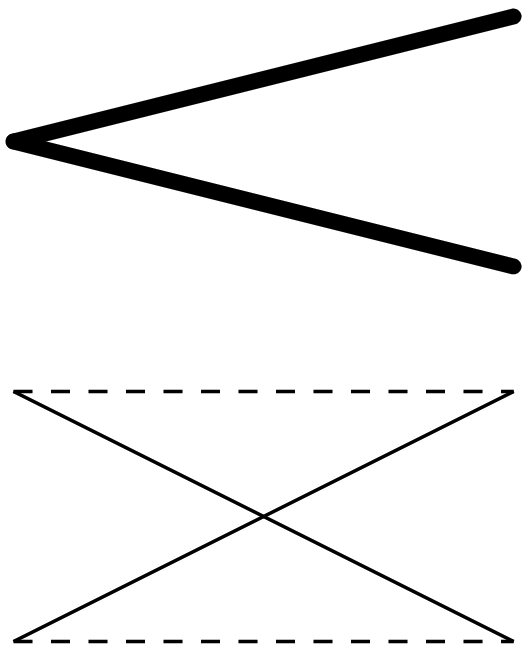} \end{tabular}}

\newcommand{\DDqd}{\begin{tabular}{c} \includegraphics[width=1cm]{diag/D42.eps} \end{tabular}}

\newcommand{\DDqt}{\begin{tabular}{c} \includegraphics[width=1cm]{diag/D43.eps} \end{tabular}}

\newcommand{\DDqq}{\begin{tabular}{c} \includegraphics[width=1cm]{diag/D44.eps} \end{tabular}}
